\documentclass{article}[12pt,a4paper]

\usepackage{geometry}
\date{}

\usepackage{setspace}
\usepackage{hyphenat} 
\usepackage[latin1]{inputenc} 
\usepackage{graphicx}
\usepackage[usenames]{color}
\usepackage{multirow}
\usepackage[FIGBOTCAP]{subfigure}
\usepackage{hyphenat} 
\usepackage[latin1]{inputenc} 
\usepackage{graphicx}
\usepackage[usenames]{color}
\usepackage{amssymb,latexsym,amsmath,srcltx,amsfonts,color}
\usepackage[latin1]{inputenc}
\usepackage{accents}
\usepackage{undertilde}
\usepackage{multirow}
\usepackage[FIGBOTCAP]{subfigure}
\usepackage{empheq,amsmath} 
\usepackage{subeqnarray}

\DeclareGraphicsExtensions{.pdf,.png,.jpg,.tif,.ps,.eps}

\usepackage[pdftex,bookmarks=true]{hyperref}

\newcommand{\bftheta}{\mbox{\boldmath $\theta$}}
\newcommand{\bfTheta}{\mbox{\boldmath $\Theta$}}

\newcommand{\bflambda}{\mbox{\boldmath $\lambda$}}

\newcommand{\bfepsilon}{\mbox{\boldmath $\epsilon$}}

\newcommand{\bfOmega}{\mbox{\boldmath $\Omega$}}

\newcommand{\bfphi}{\mbox{\boldmath $\phi$}}

\begin{document}
\titlepage
\setcounter{page}{1}
\title{\bf A mixture model for rare and clustered populations under adaptive cluster sampling}
\author{{Kelly Cristina M. Gonçalves\footnote{Departamento de Estatística, Universidade Federal Fluminense
(UFF), RJ, Brazil. Email: kelly@est.uff.br}  and Fernando A. S. Moura\footnote{Departamento de Estatística, Universidade Federal do Rio de Janeiro
(UFRJ), RJ, Brazil. Email: fmoura@dme.ufrj.br}} }
\maketitle \vspace*{-30pt}
\begin{center}
\end{center}

\begin{abstract}
  Rare populations, such as endangered species, drug users and individuals infected by rare diseases, tend to cluster in regions. Adaptive cluster designs are generally applied to obtain information from clustered and sparse populations. The aim of this work is to propose a unit-level mixture model for clustered and sparse populations when the data are obtained from an adaptive cluster sample. Our approach considers heterogeneity among units belonging to different clusters. The proposed model is evaluated using simulated data and a real experiment in which adaptive samples were drawn from an enumeration of a waterfowl species in a 5,000 $\mbox{km}^2$ area of central Florida.


\end{abstract}

\section{Introduction}\label{intro}

In many research studies, it is difficult to observe individuals or collect information from them, such as in surveys of rare diseases, elusive individuals or unevenly distributed individuals. According to \cite{mcdonald2004sampling}, rare populations present a few individuals that are sparsely distributed in clusters across a large region. In those cases, the use of conventional sampling methods is not recommended due to the high costs of locating such individuals and the low precision achieved by employing design-based estimators. For instance, suppose that the individuals of interest are spatially distributed in a region upon which we superimpose a regular grid with $N$ cells. Let $Y_i$ denote the grid cell count-for example, the number of endangered plants or animals of interest in the $i^{th}$ grid cell, $i = 1,\dots,N$. The objective is to estimate the population total $T=\sum_{i=1}^N{Y_i}$. Grid cell sampling methods involve the selection of a subset with $n<N$ grid cells and the observation of the $Y_i$'s for the selected grid cells. For rare and clustered populations, most of the samples would consist mainly of empty grid cells, yielding poor estimates of $T$.

To overcome this difficulty, \cite{thompson_adaptive_cluster_1990} introduced adaptive cluster sampling as a refined method for estimating the size of rare and clustered populations. The scheme is useful for exploring such populations because it allows sampling effort to be focused on the neighborhood of non-empty grid cells in the sample. As stated in \cite{thompson_adaptive_1996} adaptive sampling refers to designs in which the procedure for selecting units to include in the sample may depend on values of the variable of interest observed during the survey. For instance, in a survey to assess the abundance of a rare animal species, neighboring sites may be added to the sample whenever the species is encountered during the survey.

Adaptive sampling design starts with an initial probability sample of units, which is selected by a current sample design.   Then, when it has found a non-empty grid cell, it also surveys the neighbors of that cell and continues to survey neighbors of non-empty cells until it obtains a set of contiguous non-empty grid cells surrounded by empty grid cells.  Selected empty grid cells attract no additional survey effort. This procedure allows the collection of more useful data than simpler sampling methods that ignore the population structure. However, to be effective at a moderate cost, this plan requires some prior knowledge about the structure of the underlying population; see \cite{thompson_adaptive_1996} for further details.

For the particular case when the initial sample is a simple random sampling without replacement, \cite{thompson_adaptive_cluster_1990} derived inclusion probabilities for the networks observed in the sample and used these probabilities to construct design-unbiased estimators of $T$ and their variances. \cite{thompson_adaptive_cluster_1990} refers to the sets of contiguous non-empty grid cells and their neighboring empty grid cells as clusters. The set of contiguous non-empty grid cells within a cluster is called a network. Empty cells are also defined as networks of size one. The insight in \cite{thompson_adaptive_cluster_1990} was to base the analysis on networks and to treat the empty edge units of the clusters as unobserved. Adaptive cluster sampling has been performed on real problems and has been shown to be more efficient than traditional grid cell sampling in different areas. For example, \cite{roesch1993adaptive} and \cite{philippi2005adaptive} showed that this method is a viable alternative for sampling forests with rare plants. \cite{smith1995efficiency} evaluated the methodology for rare species of waterfowl, and \cite{conners2002use} applied it to hydroacoustic surveys in fisheries.

  The first attempt to model data obtained by adaptive cluster sampling and to develop a model-based Bayesian analysis was provided by \cite{rapley2008model}. The use of the Bayesian framework is a natural extension of the key idea behind adaptive cluster sampling, which incorporates the prior knowledge of a clustered population into the inference, as well as into the sampling design. The approach of \cite{rapley2008model} is based on modeling at the network level. They developed a model for the network counts that considers the informativeness of the adaptive cluster sampling design with respect to the number of counts. However, a crucial aspect of their approach is that, although they do not model the spatial locations of the networks, this decision does not entail any loss of information about the total population because, under the model, the population size does not depend on where the networks are located. They thereby address a potentially difficult problem and are able to proceed relatively simply.

Although the formulation by \cite{rapley2008model} has certain practical advantages, it does not permit the incorporation of more complex structures, such as spatial dependence between units.  Their model supposes homogeneity between all units, even units belonging to different networks, which is equivalent to assuming that the expected total in a network is proportional to its size. However, these assumptions might not be realistic in all real situations.

The aim of this work is to propose a unit-level mixture model for clustered and sparse populations when the data are obtained from an adaptive cluster sample. Our proposed mixture model considers heterogeneity among units belonging to different clusters.

The paper is organized as follows. Section \ref{model} presents the proposed model for estimating the population total of rare and clustered populations from samples selected using adaptive cluster sampling design. It also discusses prior distributions that may be used in this case. The inference specially built for fitting the proposed model is discussed in Section \ref{sec:inf}, where we also assess the convergence of the MCMC chains by applying informal and formal convergence criteria.  Section \ref{simul_study} presents a simulation study for assessing the estimation of model parameters under different scenarios. It also presents a prior sensitivity analysis of the two possible prior distributions of the parameter that controls the degree of homogeneity among units belonging to different clusters. A comparison of our approach with the one proposed by \cite{rapley2008model} through design-based and model-based perspectives under different scenarios is presented in Section \ref{sec:evaluation}. Finally, Section \ref{sec:concl} presents some conclusions and suggestions for further research.

\section{A Poisson mixture model for unit counts}\label{model}

The basic mixture model for independent scalar or vector observations $Y_i$, $i=1,...,n$ is given by:
\begin{eqnarray}\label{mixt}
Y_i\sim \sum_{j=1}^k{w_jf(\cdot\mid \phi_j)},\, i=1,\dots,n,
\end{eqnarray}
where $f(\cdot\mid \bfphi)$ is a given parametric family of densities indexed by a scalar or a vector $\bfphi$. In general, the objective of the analysis is to make inferences about the unknowns: the number of groups, $k$; the parameters $\phi_j$'s and the components' weights, $w_j$, $0<w_j<1$, $\sum_{j=1}^k{w_j}=1$.  The mixture model in (\ref{mixt}) is invariant to permutation of the labels $j=1,\dots,k$. Therefore, it is important to adopt unique labeling to ensure identifiability. For example, we can impose an ordering constraint on $\phi_j$'s, such as $\phi_1<\phi_2<\dots<\phi_k$.

\cite{viallefont2002bayesian} suggest a Poisson mixture model for dealing with rare events.
The interest in this class of models arises here, because it is applicable to heterogeneous populations consisting of groups $j=1,\dots,k$ of sizes proportional to $w_j$, from which a random sample may be drawn. The identity of the group from which each observation is drawn is unknown. As stated in \cite{richardson1997bayesian}, due to computational costs, it is natural to regard the group label $\epsilon_i$, for the $i$-th observation as a latent variable and rewrite (\ref{mixt}) as the following  hierarchical model:
\begin{eqnarray*}
Y_i\mid \phi_j,\epsilon_i=j\sim f(\cdot\mid \phi_{j}), \mbox{ with } P(\epsilon_i=j)=w_j,\, i=1,\dots,n,\, j=1,\dots,k.
\end{eqnarray*}

Let us consider a region $\bfOmega$ containing a sparse, clustered population of size $T$. We superimpose a regular grid on $\bfOmega$ to partition it into $N$ squares. A grid cell is non-empty if it contains at least one observation and empty otherwise. Let $X$ be the number of non-empty grid cells in $\bfOmega$. Let $R\leq X$ be the number of non-empty networks, and let ${\bf C}=(C_1,\dots,C_R)'$ denote the number of non-empty grid cells within each network, so that $X=\sum_{j=1}^R{C_j}$. As there are $N-X$ empty grid cells, which are defined to be empty networks of size one, there are $N-X+R$ networks in $\bfOmega$. Thus, it is possible to extend the $R$-vector ${\bf C}$ to the vector ${\bf Z}=({\bf C}',{\bf 1}_{N-X}')'$ of dimension $N-X+R$, where ${\bf 1}_{N-X}'$ is the vector of ones with dimension $N-X$. Let ${\bf Y}=(Y_1,\dots,Y_{X})'$ denote the vector of cell counts, where its elements are the number of observations within each non-empty unit; then, $Y_i\geq 1$. The main goal is to make inferences about the total population $T=\sum_{i=1}^X{Y_i}$.

The proposed mixture model assumes that the $R$ non-empty network mixture components are heterogeneous, with weights $w_j$, which in each case are proportional to the number of grid cells inside the networks, $C_j$.  Let us define the latent allocation variable $\epsilon_i$ such that $P(\epsilon_i = j) = w_j = C_j/X$, $i=1,\dots,X$ and $j=1,\dots,R$.

 The mixture model is completed with the hierarchical structure proposed in \cite{rapley2008model}, where they assign distributions to $X$, $R$ and ${\bf C}$ associated with the non-empty grid cells and then, conditionally on the network structure, model the network counts ${\bf Y}$ for the non-empty networks.

Our proposed model, can be stated as follows:
\begin{subequations}\label{model_mixt_prop}
\begin{eqnarray}
Y_i\mid \epsilon_i=j,\lambda_j, X &\sim &\mbox{ independent truncated Poisson}(\lambda_{j}), \; Y_i\geq 1,\label{proposal1}\\
P(\epsilon_i = j) &= &w_j = C_j/X,\; i=1,\dots,X \mbox{ and } j=1,\dots,R,\label{proposal2}\\
{\bf C}\mid X, R &\sim &{\bf 1}_R + \mbox{ Multinomial }\left(X-R,\frac{1}{R}{\bf 1}_R\right),\; \sum_{i=1}^R{C_i}=X,\label{rapley1}\\
R\mid X,\beta &\sim &\mbox{ truncated Binomial }(X,\beta),\; R=1,\dots,X,\label{rapley2}\\
X\mid \alpha &\sim &\mbox{ truncated Binomial }(N,\alpha),\;X=1,\dots,N,\label{rapley3}
\end{eqnarray}
\end{subequations}
where $\lambda_j/\{1-\exp(-\lambda_j)\}$ is the mean of the truncated Poisson distribution, and  ${\bf 1}_R$ is the R-vector of ones. It should be noted that, to avoid degeneracy, there is assumed to be at least one non-empty network in the region. Consequently, all the distributions are left-truncated at one.

The distributions stated in (\ref{rapley1}), (\ref{rapley2}) and (\ref{rapley3}) are the same as in the model by \cite{rapley2008model}, but unlike their model, the analysis here is performed at the unit level.  In the \cite{rapley2008model} model, the equations (\ref{proposal1}) and (\ref{proposal2}) are replaced with independent Poisson distributions truncated at zero: $Y_{.j}\mid \lambda, R, {\bf C} \sim \mbox{ Poisson }(\lambda C_j)$, where $Y_{.j}=\sum_{i\in U_j}{Y_i}$ with $U_j$ denoting the set of units that belong to the network $j, j=1,...,R$.  Therefore, our model can handle heterogeneity between units that belong to different networks, which is not considered in the approached proposed by  \cite{rapley2008model}.

The selection mechanism that leads to a particular sample $s=\{i_1,\dots,i_m\}$ of size $m$ taken from $N-X+R$ networks is also included in the model and depends only on the network structure, described by $X, R$ and ${\bf C}$. We consider sampling designs whose networks are sampled directly via a sequential procedure where the ordered sample of networks is selected without replacement.

Networks are sampled by the method by probability proportional to size without replacement. Note that the inclusion probability of a network depends on its size $Z_i$, and the sampling is informative because the components of the random vector ${\bf Z}$ are only observed for the sampled networks after being selected. Thus, the probability of selecting the ordered sample $s=\{i_1,\dots,i_m\}$ of $m$ networks must be included in the model likelihood. The joint inclusion probability can be deduced as follows.

Let the event $A_{i_j} = \{\mbox{the network } i_j \mbox{ be selected in the } j\mbox{-th } \mbox{ draw}\}.$ Thus, the probability of selecting the ordered sample $s=\{i_1,\dots,i_m\}$ of $m$ networks can be written as follows:
\begin{align} \label{ps}
\begin{array}{ll}
p(s\mid X, R, {\bf C})= P(\cap_{j=1}^m A_{i_j}\mid X, R, {\bf C}) &= P(A_{i_1} \mid X, R, {\bf C})\\
&\times\prod_{j=2}^{m} P(A_{i_j} \mid  \cap_{k=1}^{j-1} A_{i_k}, X, R, {\bf C}).
\end{array}
\end{align}
Because the networks are sampled without replacement, the conditional probabilities $P(A_{i_1} \mid X, R, {\bf C})$ and $P(A_{i_j} \mid \cap_{k=1}^{j-1} A_{i_k}, X, R, {\bf C})$ in (\ref{ps}) are, respectively, given by:
\begin{align} \label{pi}
\begin{array}{rl}
P(A_{i_1} \mid X, R, {\bf C})&= \frac{z_{i_1}\times g_{i_1,1}}{\sum_{i=1}^{N-X+R}{z_i}- {z_{i_0}}} \\
P(A_{i_j} \mid \cap_{k=1}^{j-1} A_{i_k}, X, R, {\bf C})&= \frac{z_{i_j}\times g_{i_j,j}}{\sum_{i=1}^{N-X+R}{z_i}-\sum_{k=0}^{j-1}{z_{i_k}}}, j=2,\dots,m,
\end{array}
\end{align}
where $g_{i_j,j}$ is the number of unselected networks of size $z_{i_j}$  after $j-1$ networks have been selected and $z_{i_0}=0$.

Substituting the equations in (\ref{pi}) into (\ref{ps}), we finally have:
\begin{eqnarray}\label{prob_sel_mixt}
p(s\mid X, R, {\bf C}) = \prod_{j=1}^m{\frac{z_{i_j}\times g_{i_j,j}}{\sum_{i=1}^{N-X+R}{z_i}-\sum_{k=0}^{j-1}{z_{i_k}}}}.
\end{eqnarray}
The sampling procedure entails observing $Y_i$ for the networks in the sample $s$. The input variables are split  into an observed component and an unobserved one, using the subscripts $s$ and $\bar{s}$, respectively. Thus, we have $X=X_s+X_{\bar{s}}$, $R=R_s+R_{\bar{s}}$, $\bfepsilon=(\bfepsilon_s',\bfepsilon_{\bar{s}}')'$, ${\bf C}=({\bf C}_s',{\bf C}_{\bar{s}}')'$  and ${\bf Y}=({\bf Y}_s',{\bf Y}_{\bar{s}}')'$.

As the sampling procedure is informative, it is useful to break the joint probability model into two parts: the model for the underlying complete data, including both observed and unobserved components, and the model for the inclusion probability vector, as stated in (\ref{prob_sel_mixt}) (see \cite{pfeffermann2006multi} for further explanation). The complete-data likelihood is defined as the product of these two factors, as stated by \cite{gelman1995bayesian}. Thus, we can write the complete-data likelihood as
\begin{eqnarray}\label{compl_likelihood}
\nonumber &p(\{i_1,\dots,i_m\}, X, R, {\bfepsilon},{\bf C},{\bf Y}\mid\bflambda, \alpha, \beta)= p(\{i_1,\dots,i_m\}\mid X, R, {\bf C})p({\bf Y}\mid \bfepsilon,\bflambda, X)\\
\nonumber & \times p(\bfepsilon\mid {\bf C}, R, X) p({\bf C}\mid R,X) p(R\mid X, \beta)p(X\mid \alpha)\\
&=\displaystyle\prod_{l=1}^m{\frac{z_{i_l}\times g_{i_l,l}}{\sum_{i=1}^{N-X+R}{z_i}-\sum_{k=0}^{j-1}{z_{i_k}}}}\times\prod_{j=1}^{R_s+R_{\bar{s}}}{\prod_{\{i:\epsilon_i=j\}}{\frac{\lambda_j^{y_i}\exp(-\lambda_j)}{y_i![1-\exp(-\lambda_j)]}}}\\
\nonumber &\times\frac{1}{{(X_s+X_{\bar{s}})^{X_s+X_{\bar{s}}}}}\prod_{j=1}^{R_s+R_{\bar{s}}}{C_j^{C_j}}\times\prod_{j=1}^{R_s+R_{\bar{s}}}\frac{1}{(C_j-1)!}\left(\frac{1}{R_s+R_{\bar{s}}}\right)^{C_j-1}\\
\nonumber &\frac{1}{(R_s+R_{\bar{s}})!}\frac{\beta^{R_s+R_{\bar{s}}}(1-\beta)^{X_s+X_{\bar{s}}-R_s-R_{\bar{s}}}}{1-(1-\beta)^{X_s+X_{\bar{s}}}}\times N!\frac{\alpha^{X_s+X_{\bar{s}}}(1-\alpha)^{N-X_s-X_{\bar{s}}}}{1-(1-\alpha)^N}.
\end{eqnarray}

It should be noted that  expression (\ref{compl_likelihood}) is useful for setting up a probability model, but it is not actually the likelihood of the data unless the variables are completely observed. The appropriate likelihood of Bayesian inference for the actual information available is obtained by summing over the unknown quantities and not otherwise observed in the selected sample.  The observed-data likelihood, conditional on $\bflambda$, $\alpha$ and $\beta$, is given by:
\begin{eqnarray*}
p(\{i_1,\dots,i_m\},X_s, R_s, {\bfepsilon}_s, {\bf C}_s,{\bf Y}_s)= \sum_{{\bf Y}_{\bar{s}}}\sum_{{\bf C}_{\bar{s}}}\sum_{{\bfepsilon}_{\bar{s}}}\sum_{R_{\bar{s}}}\sum_{X_{\bar{s}}}{p(\{i_1,\dots,i_m\}, X, R, \bfepsilon, {\bf C},{\bf Y})}.
\end{eqnarray*}

\subsection{Prior distributions}\label{sec:prior}

\begin{figure}[h!]
\hspace{-0.8 cm}\begin{tabular}{l}
\subfigure[$(\alpha,\beta)=(0.05,0.05)$]{\includegraphics[scale=0.34]{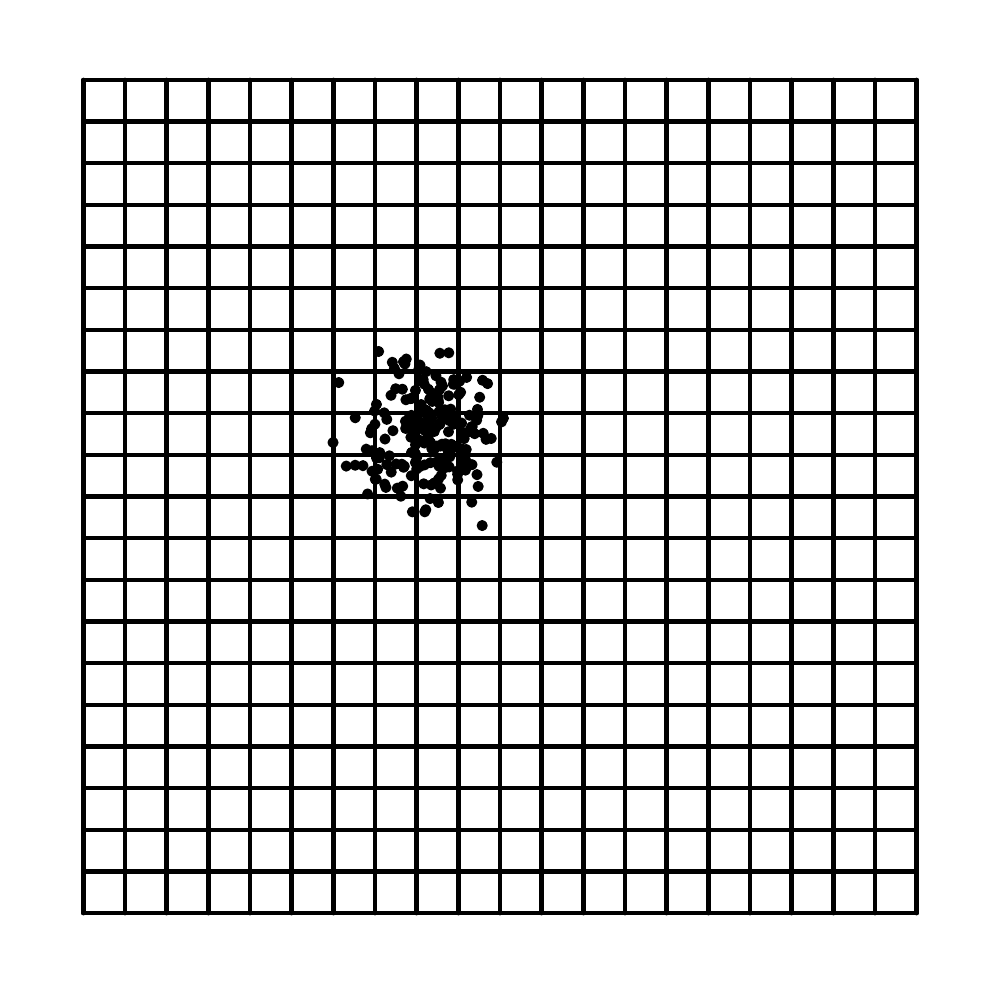}}
\subfigure[$(\alpha,\beta)=(0.05,0.2)$]{\includegraphics[scale=0.34]{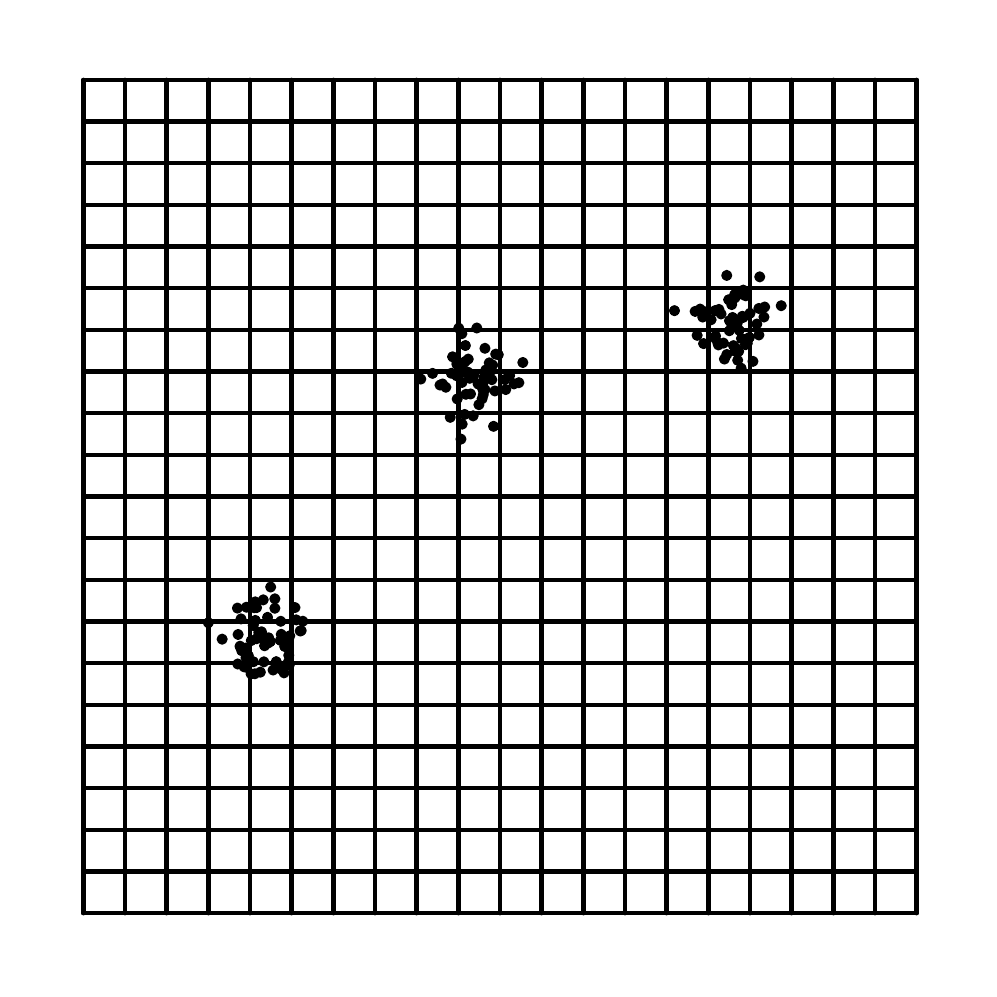}}
\subfigure[$(\alpha,\beta)=(0.2,0.05)$]{\includegraphics[scale=0.34]{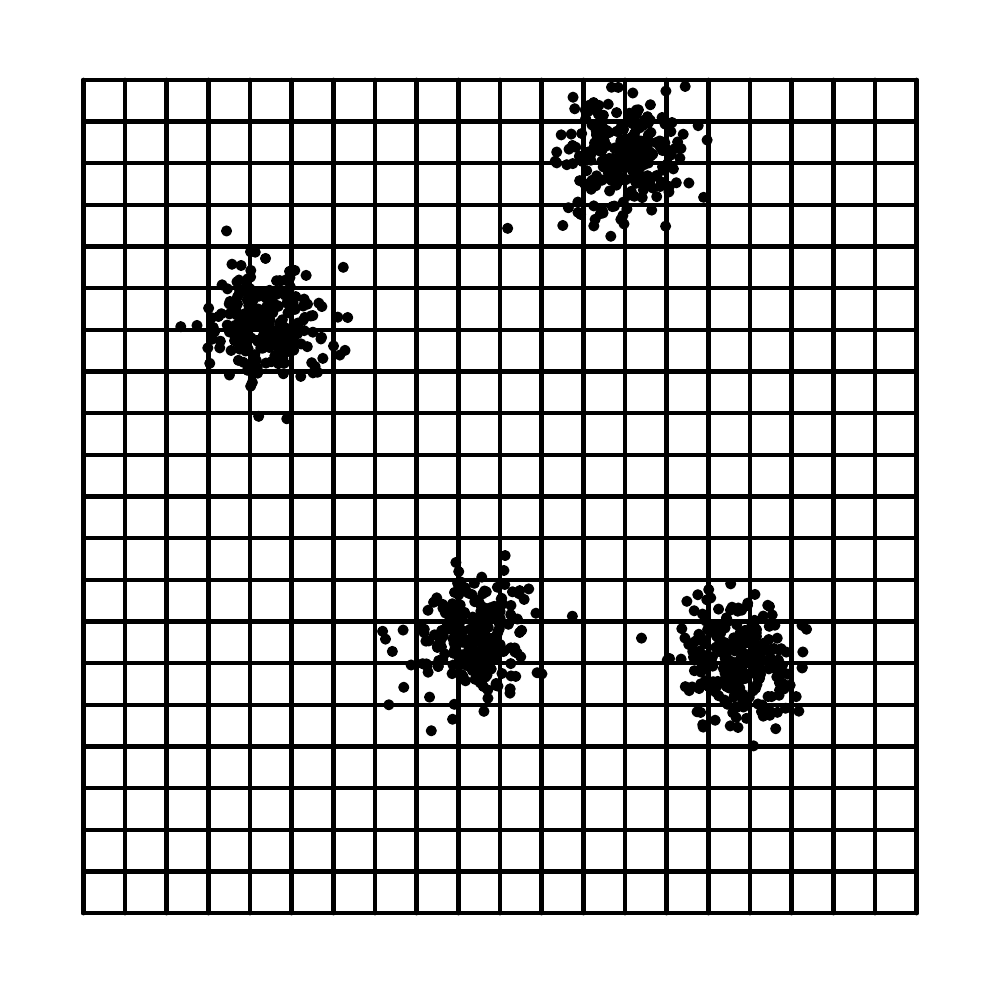}}
\subfigure[$(\alpha,\beta)=(0.2,0.2)$]{\includegraphics[scale=0.34]{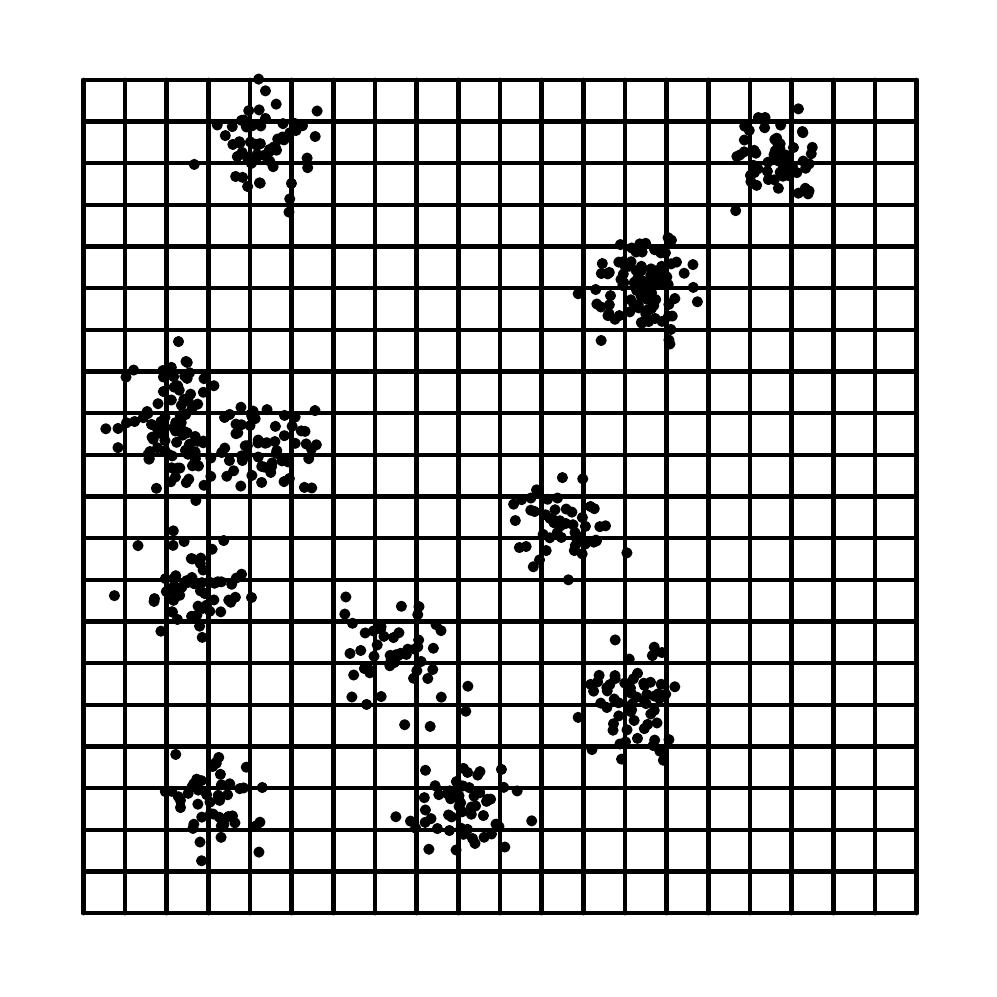}}
\end{tabular}
\vspace{-0.4 cm}\caption{{\it Artificial populations generated by the proposed model with some fixed values for $\alpha$ and $\beta$, and $\lambda_j = 10$, for all $j=1,\dots,R$, in a regular grid with $N=400$ units.}}\label{cenarios}
\end{figure}

 In a Bayesian framework, the three unknowns $\alpha$, $\beta$ and $\bflambda$ are regarded as having been drawn from appropriate prior distributions. Assume that these parameters are independent; then, the joint prior distribution of $(\alpha,\beta,\bflambda)$ is the product of their marginal prior distributions, described here. The parameter $\alpha$ controls the expected number of non-empty grid cells, and $\beta$ controls the conditional expected number of non-empty networks. Figure \ref{cenarios} presents an illustration with certain artificial populations generated by model (\ref{model_mixt_prop}) and certain values fixed for $\alpha$ and $\beta$. We also arbitrarily fixed $\lambda_j = 10$, for all $j=1,\dots,R$; thus, approximately $10$ observations are expected in each unit.
\hspace{-1.5 cm}

Because our approach aims to survey sparse populations, when analyzing Figure \ref{cenarios}, it is reasonable to assume that both the $\alpha$ and $\beta$ parameters should typically be small. To be uninformative with respect to these parameters, we should choose flat prior distributions.  However, we can assign prior distributions that incorporate our knowledge of a rare and clustered population.  In particular, we can consider that $\alpha\sim \mbox{ Beta } (a_\alpha,b_\alpha)$ and $\beta\sim \mbox{ Beta } (a_\beta,b_\beta)$ and choose values for the Beta distribution's parameters such that $\alpha$ and $\beta$ are within an interval centered on a small value with high probability. The symbol $W\sim \mbox{Beta}(a,b)$ generically denotes that $W$ is beta distributed and parameterized with mean $a/(a+b)$ and variance $ab(a+b+1)^{-1}(a+b)^{-2}$.

To ensure identifiability, it is necessary to adopt a unique labeling. For the proposed model in (\ref{model_mixt_prop}), unique labeling can be achieved by imposing a restriction on $\bflambda= (\lambda_1,\dots,\lambda_R)'$. However, it should be noted that $\bflambda$, although totally unknown, has components associated with the sample where better estimates are expected. Thus, let us define $\bflambda=(\bflambda_s,\bflambda_{\bar{s}})'$, such that $\bflambda_s$ refers to the networks observed in the sample and $\bflambda_{\bar{s}}$ to the unobserved networks. Note that it is necessary to impose a restriction on $\bflambda$ to ensure the identifiability of the model. Nevertheless, this restriction is only necessary  for the elements of $\bflambda$ associated with the unknown networks, i.e., $\bflambda_{\bar{s}}$.

   Let us assume the following for $\bflambda$:
\begin{align*}\label{prior}
{\bflambda}\mid \bftheta &\sim p(.\mid \bftheta,R),\, \mbox{ such that } \lambda_j<\lambda_{j+1}, \mbox{ for all } j\in [R_s+1,R_s+R_{\bar{s}}),
\end{align*}
where $p(.\mid \bftheta,R)$ represents the prior distribution of $\bflambda$, which depends on the number of networks in the population, $R$, and on the vector of hyperparameters $\bftheta$.

We use two different prior distributions for $\bflambda$. First, we assume that the $\lambda_j$'s are  conditionally independent given $\bftheta$ each with prior density $p(\lambda_j\mid \bftheta)$. Then, the joint prior density for $\bflambda$ is given by the following:
$$p(\bflambda\mid\bftheta,R)=R_{\bar{s}}!p(\lambda_1\mid\bftheta)\dots p(\lambda_R\mid \bftheta),\mbox{ such that } \lambda_j<\lambda_{j+1}, \mbox{ for all } j\in [R_s+1,R_{\bar{s}}).$$
In particular, we consider $\lambda_j\sim\mbox{Gamma}(d,\nu)$, $\bftheta=(d,\nu)$ and introduce an additional hierarchical level by allowing $\nu$ to follow a $\mbox{Gamma}(e,f)$. The symbol $W\sim \mbox{Gamma}(a,b)$ generically denotes that $W$ is gamma distributed and parameterized with mean $a/b$ and variance $a/b^2$.

One standard way of setting a Gamma as a weakly informative prior is to choose small values for its two parameters. However, such a distribution has a peak in the neighborhood of zero, which might encourage the inclusion of components with very small Poisson parameters, which would be difficult to estimate in general. Therefore, we used a weakly informative prior based on \cite{viallefont2002bayesian}-i.e., $\mbox{Gamma}(d,\nu)$ with  $d$ greater than one-to avoid the exponential shape without overly reducing the coefficient of variation's (CV)  distribution. The parameter $\nu$ is set such that the prior mean $d/\nu$ is equal to the midrange of the observed data. However, in our case, we also consider $\nu$ to be unknown, so we choose $e$ and $f$ in the prior of $\nu$ such that the approximation to the mean of $\lambda_j$, $d/(e/f)$, is equal to the midrange of the observed data and the variance $e/f^2$ is relatively small.

The other prior considered for $\bflambda$ is the one introduced by \cite{roeder1997} for normal mixtures as an explicit way to place an informative prior on the distance between two consecutive $\lambda_j$'s. Here, the hyperparameter $\bftheta$ is $\tau$, a positive constant, and the prior model is given by the following:
$$p(\bflambda\mid \tau, R) = p(\lambda_R\mid\lambda_{R-1}, \tau)p(\lambda_{R-1}\mid\lambda_{R-2}, \tau)\dots p(\lambda_1),$$
where $p(\lambda_j\mid\lambda_{j-1}, \tau)$ is $\mbox{N}_{(\lambda_{j-1},\infty)}(\lambda_{j-1},\tau)$, i.e., a Normal centered at $\lambda_{j-1}$ with variance $\tau^2$, truncated to be greater than $\lambda_{j-1}$ and $p(\lambda_1)\propto 1$. This ordering ensures the identifiability of the model.

 \cite{viallefont2002bayesian} illustrate the difficulty of eliciting $\tau$ and its clear influence on the posterior distribution of the mixture parameters, as well as on the posterior distribution of the number of components. For example, if $\tau$ is very small compared to the anticipated distance between two consecutive $\lambda_j$'s, there will be a tendency to fit intermediate components between the true ones and hence to find a posterior distribution favoring higher values of $R$. This strategy gives a low prior probability that any two neighboring components are more than $\tau$ standard deviations apart.  Based on a simulation study, \cite{roeder1997} recommend choosing  $\tau=5$ because this choice leads to reasonable density estimates.

\section{Inference}\label{sec:inf}

The posterior distributions of the parametric vector $\bfTheta=(X_{\bar{s}}, R_{\bar{s}}, \bfepsilon_{\bar{s}},{\bf C}_{\bar{s}}, {\bf Y}_{\bar{s}},\alpha,\beta,\bflambda,\nu)$ of model (\ref{model_mixt_prop}) cannot be obtained in closed form. Therefore, it is necessary to use some numerical approximation methods. One alternative, which is often used and is feasible to implement, is to generate samples from the marginal distributions of the parameters based on the Markov Chain Monte Carlo (MCMC) algorithm. Nevertheless, this method, as originally formulated, requires the posterior distribution to have a density with respect to some fixed measure. Thus, it cannot be used alone in this case, where the size of the parametric space is also a parameter. We use an approach based on reversible jump MCMC (RJMCMC), which was first proposed in \cite{green1995reversible} and applied in mixture models with unknown numbers of components by \cite{richardson1997bayesian}. The method basically consists of jumps between the parameter subspaces corresponding to different numbers of components in the mixture.

For the proposed model (\ref{model_mixt_prop}), we used the steps specified below:
\begin{itemize}
\item [(1)] update the parameters $\alpha$, $\beta$, $\bftheta$ and $\bflambda$;
\item [(2)] update the unobserved variables $X_{\bar{s}}$ and ${\bf Y}_{\bar{s}}$;
\item [(3)] update the allocation $\bfepsilon_{\bar{s}}$ so that ${\bf C}_{\bar{s}}$ is updated; and
\item [(4)] combine two networks into one, or split one into two.
\end{itemize}

Steps (1)-(3) are performed using the Gibbs sampler or a Metropolis-Hastings sampler, and they do not change the dimensions of $\bfTheta$. It should be noted that, because the proposed model (\ref{model_mixt_prop}) is defined only for the non-empty units, it is not possible to update the allocation, resulting in networks without any observations. Consequently, this step needs to be restricted so that each network must have at least one observation.

Step (4) involves changing $R_{\bar{s}}$ by $1$ and making the necessary corresponding changes to $(\bflambda,{\bf C},\bfepsilon)$. We made a random choice between splitting and combining, with probabilities depending on $R_{\bar{s}}$. Let $\lambda_j'=\lambda_j/\{1-\exp(-\lambda_j)\}$ be the mean of the truncated Poisson distribution. The combination proposal begins by choosing a pair of components $(j_1,j_2)$ at random, such that $\lambda'_{j_1}<\lambda'_{j_2}$. These two components are merged, forming a new component $j^*$. Now, we have to reallocate all the observations with $\epsilon_i=j_1$ or $\epsilon_i=j_2$ and create values for $(w_{j^*},\lambda'_{j^*})$. They are chosen such that
\begin{align*}
w_{j^*}&=w_{j_1}+w_{j_2},\\
w_{j^*}\lambda'_{j^*}&=w_{j_1}\lambda'_{j_1}+w_{j_2}\lambda'_{j_2},
\end{align*}
and we must impose $\lambda'_{j-1}<\lambda'_{j_1}<\lambda'_{j_2}<\lambda'_{j+1}$.  A component $j^*$ is chosen at random and split into $j_1$ and $j_2$. However, there are two degrees of freedom for achieving this step, so we need to generate a two-dimensional random vector ${\bf u}=(u_1,u_2)$ to specify the new parameters. \cite{viallefont2002bayesian} present some ways of proposing a split that enforces the positivity constraint on Poisson parameters. In this work, we used the one referenced as ``SM2" in their paper. In particular, the proposed model (\ref{model_mixt_prop}) is applicable to non-empty networks; thus, the split proposal also requires that both networks have at least one observation. Therefore, networks with only one observation cannot be chosen to be split. The acceptance probability for the split and combination steps can be viewed in Appendix \ref{sec:appendix}.

\begin{figure}[h!]
\begin{center}
\vspace{-0.3 cm}\begin{tabular}{cc}
{\includegraphics[scale=0.5]{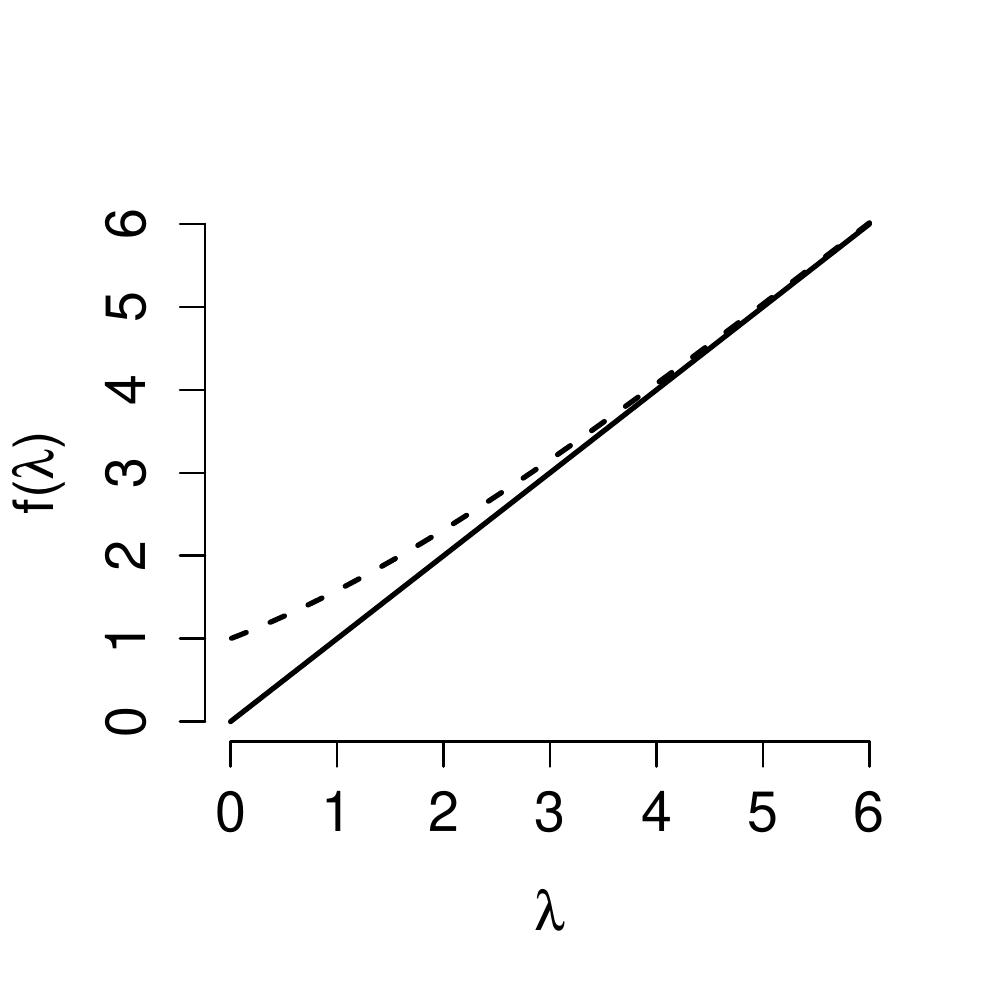}}
\end{tabular}
\end{center}
\vspace{-0.7 cm}\caption{{\it Comparison of the first-order moments of the Poisson distribution (\----) and of the Poisson distribution truncated at zero (\--\,\--\,\--).}}\label{comp}
\end{figure}

Although the expression above can be written in terms of $\lambda_j'$, the likelihood is expressed in terms of $\lambda_j$. Therefore, after step (4), we need to obtain $\lambda_j$ from $\lambda_j'$ by solving the equation $\lambda_j'=\lambda_j/\{1-\exp(-\lambda_j)\}$. Furthermore, although the target function is invertible, it involves a polynomial with an exponential function, for which, in general, it is impossible to obtain an exact analytical solution. When the value of $\lambda_j$ is sufficiently large, we can approximate $\lambda_j$ by $\lambda_j'$ (see Figure \ref{comp}). However, for cases in which this approximation is not good, we need to use a numerical approximation, such as the Taylor approximation.

\subsection{Convergence diagnostics}\label{conv}

 To assess the performance of the proposed model and to check the convergence of the RJMCMC estimation, we generated a clustered population in an area with $N=400$ units, fixing $\alpha=0.15$ and $\beta=0.10$. The values of the components of $\bflambda$  were generated from a Gamma distribution centered in $8.5$ with a coefficient of variation (CV) equal to $95\%$, resulting in a Gamma distribution with parameters $d=1.1$ and $\nu=0.13$. Then, we selected a 5\% sample using the adaptive design. We considered the prior distributions described in Section \ref{sec:prior}. For $\alpha$ and $\beta$, we chose $a_{\alpha}=3$, $a_{\beta}=15$, $b_{\alpha}=1$ and $b_{\beta}=9$, which parallel the prior distributions considered by \cite{rapley2008model}. These values are suitable when the only knowledge that can be obtained about the underlying population is that it is sparse and clustered. For $\bflambda$, we considered only the Gamma independent prior used in the generation of the artificial data. The population generated yields $R=8$ networks and, the networks observed were $s=\{2,4,7\}$, labeled such that the components of  $\bflambda$ are in increasing order.

\begin{figure}[h!]
\begin{tabular}{cccc}
\hspace{-0.5 cm}{\includegraphics[scale=0.4]{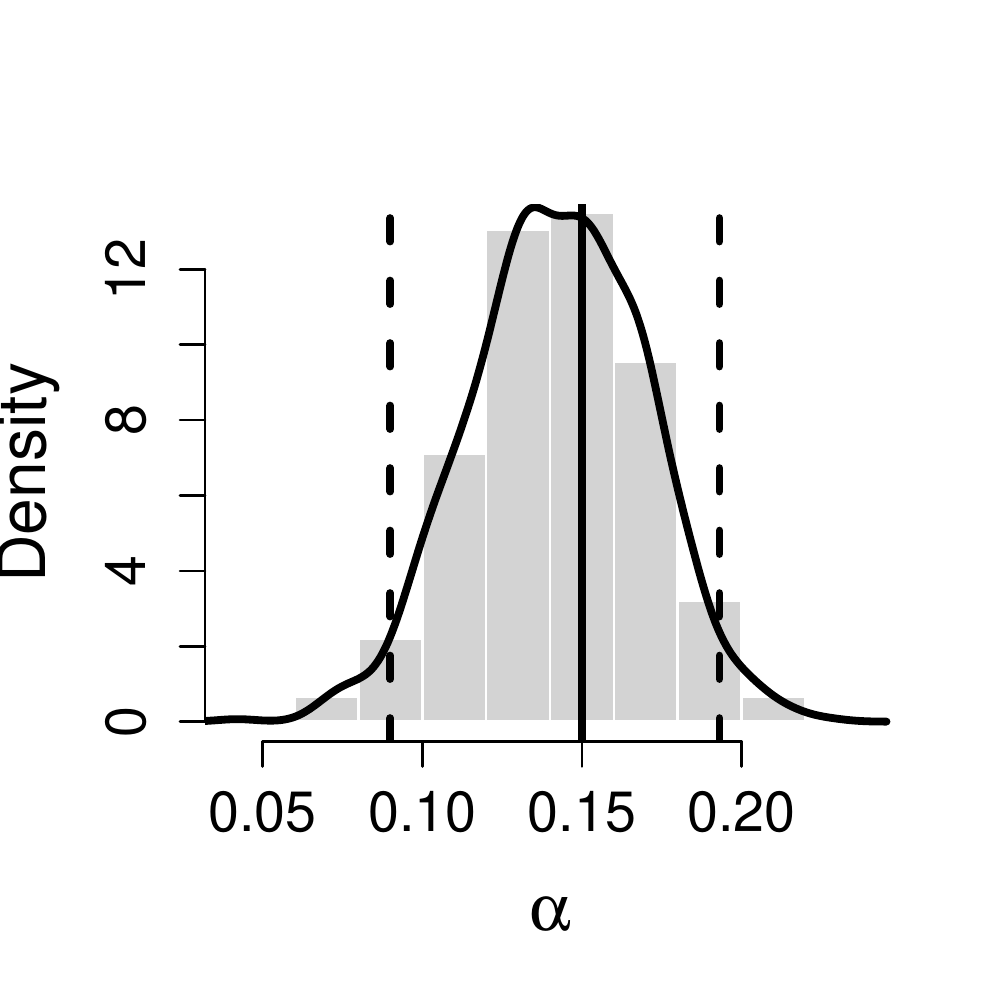}}
\hspace{-0.5 cm}{\includegraphics[scale=0.4]{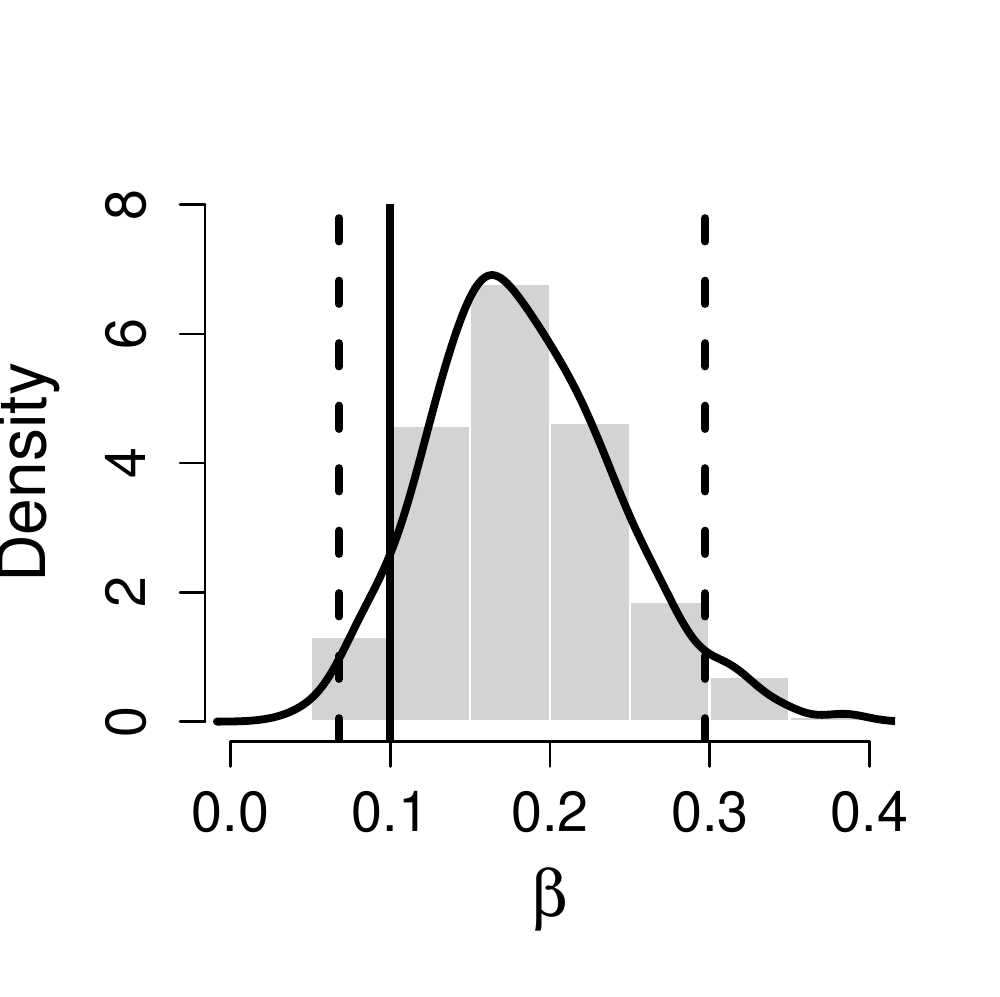}}
\hspace{-0.5 cm}{\includegraphics[scale=0.4]{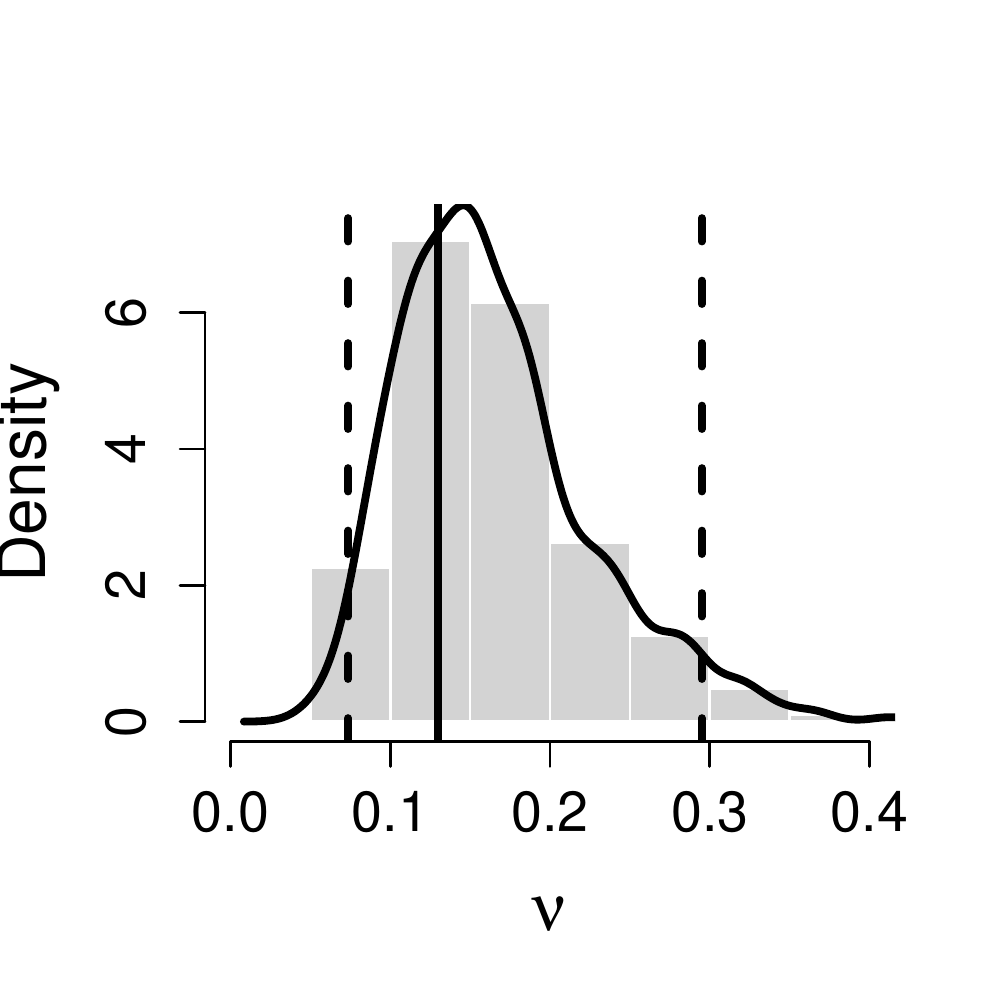}}
\hspace{-0.5 cm}{\includegraphics[scale=0.4]{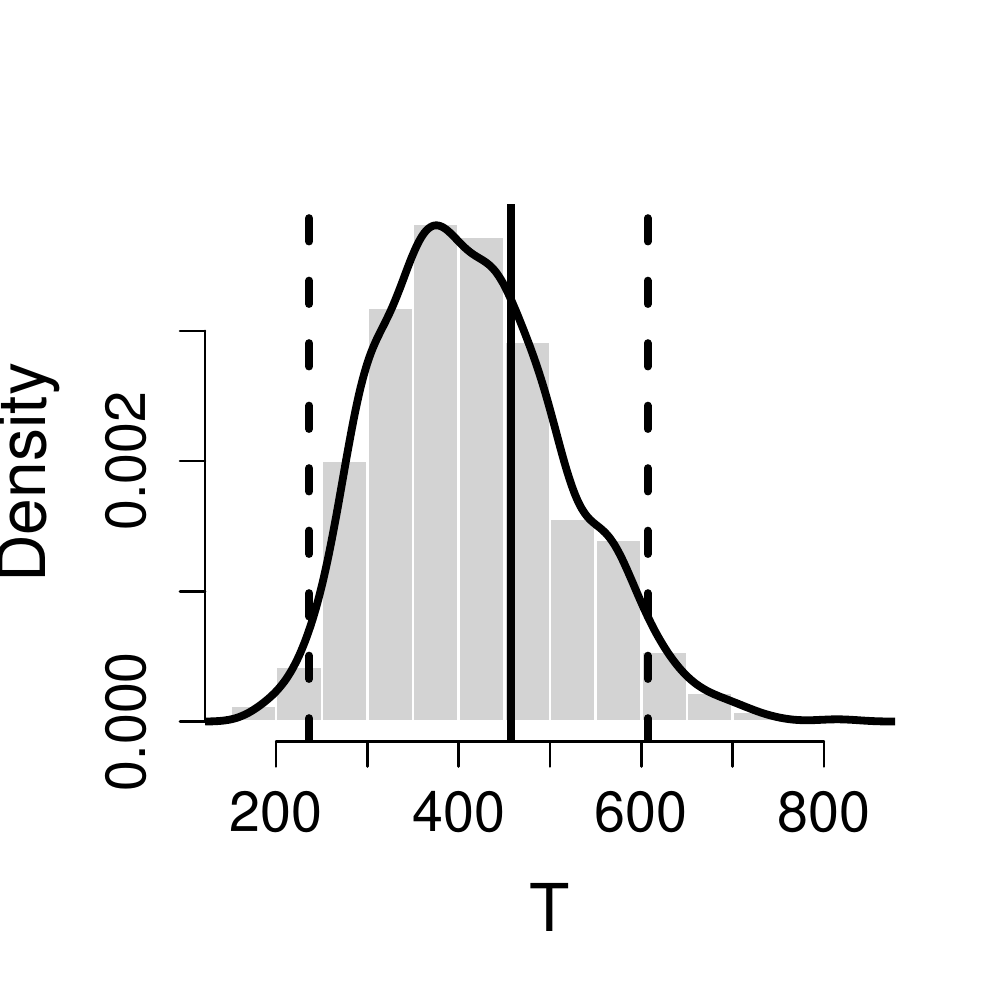}}\\
\hspace{-0.5 cm}{\includegraphics[scale=0.4]{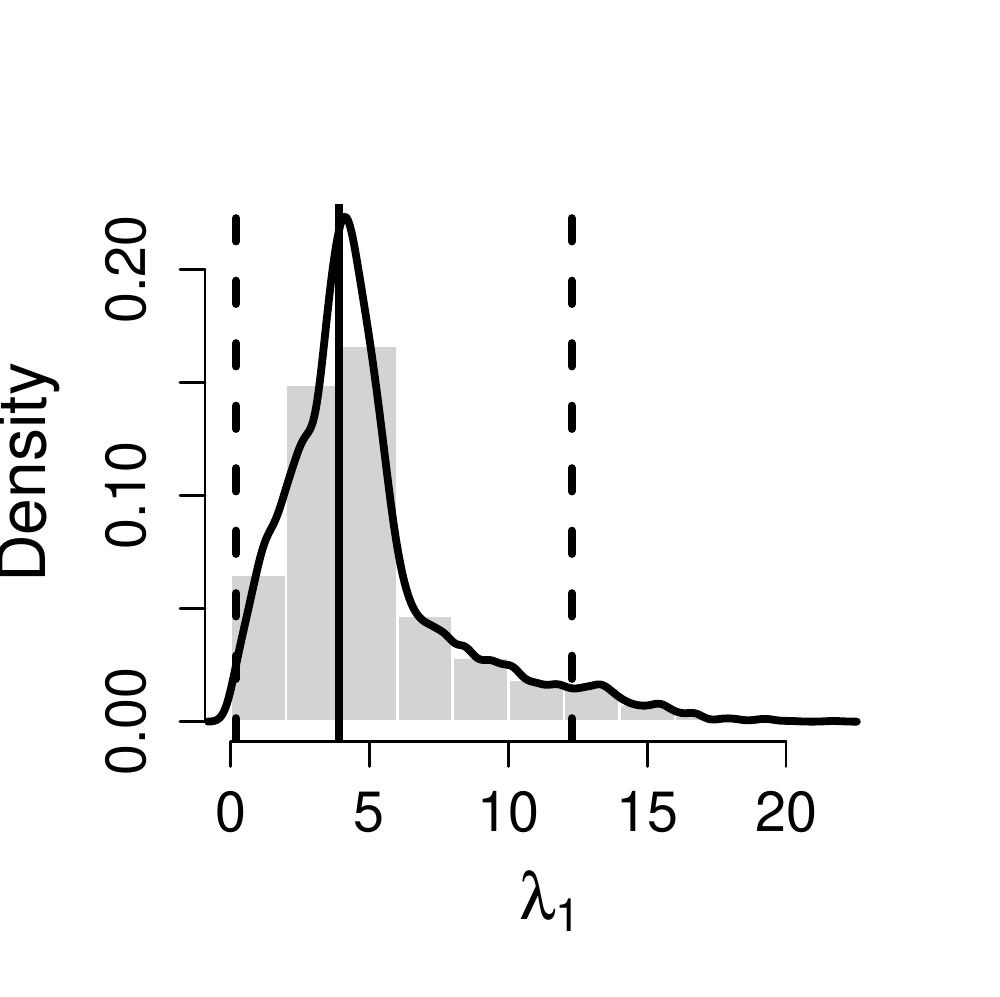}}
\hspace{-0.5 cm}{\includegraphics[scale=0.4]{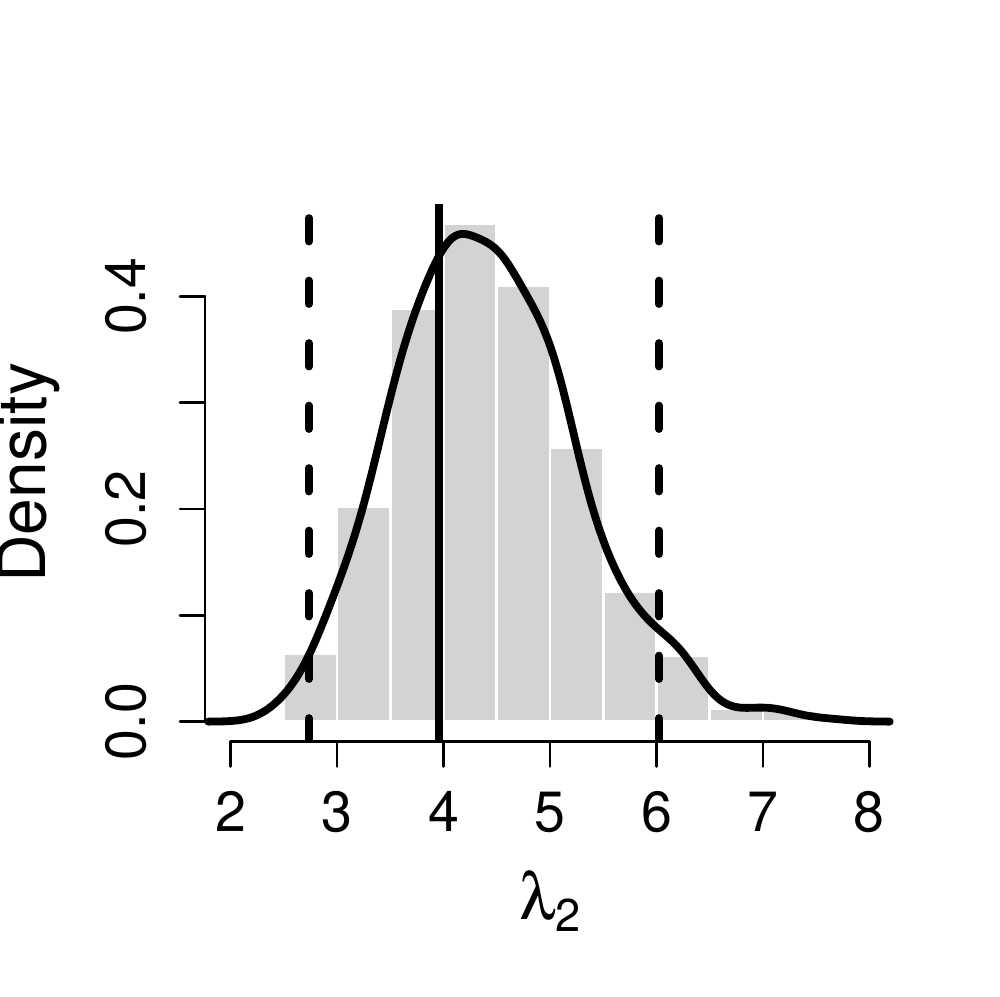}}
\hspace{-0.5 cm}{\includegraphics[scale=0.4]{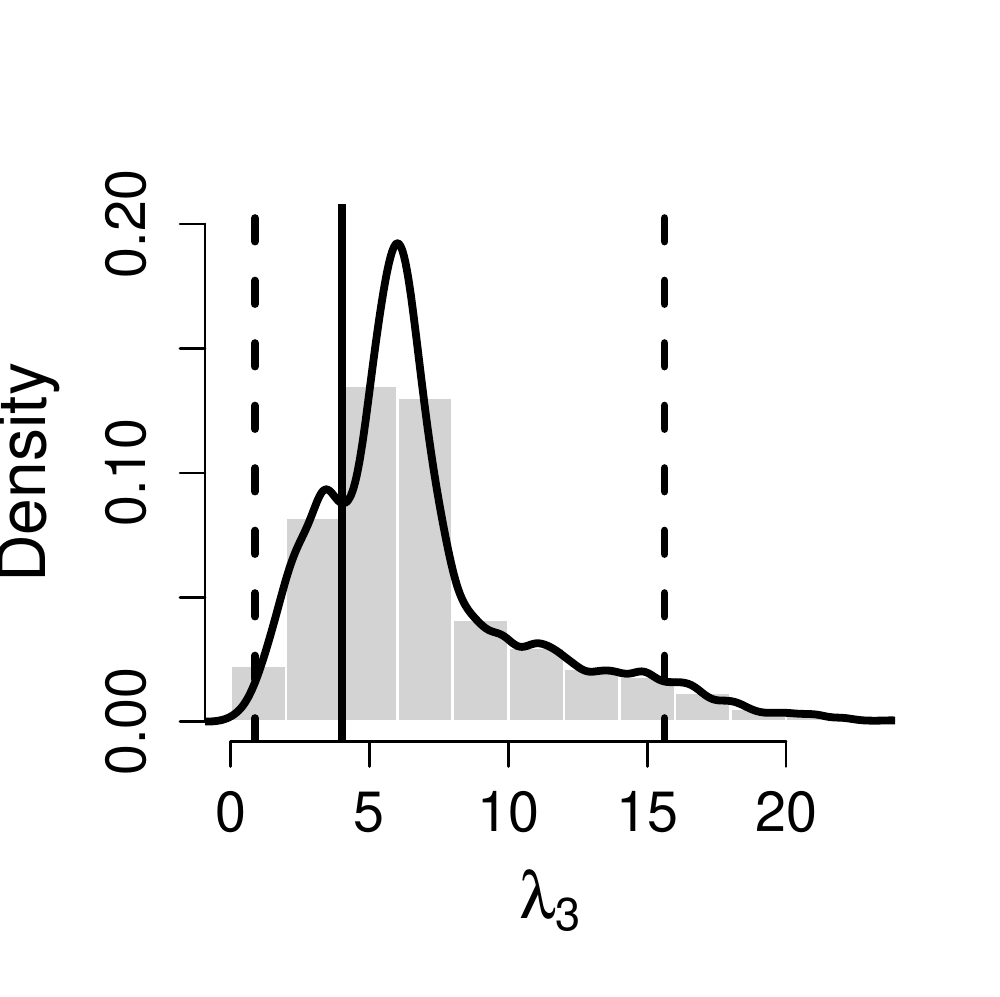}}
\hspace{-0.5 cm}{\includegraphics[scale=0.4]{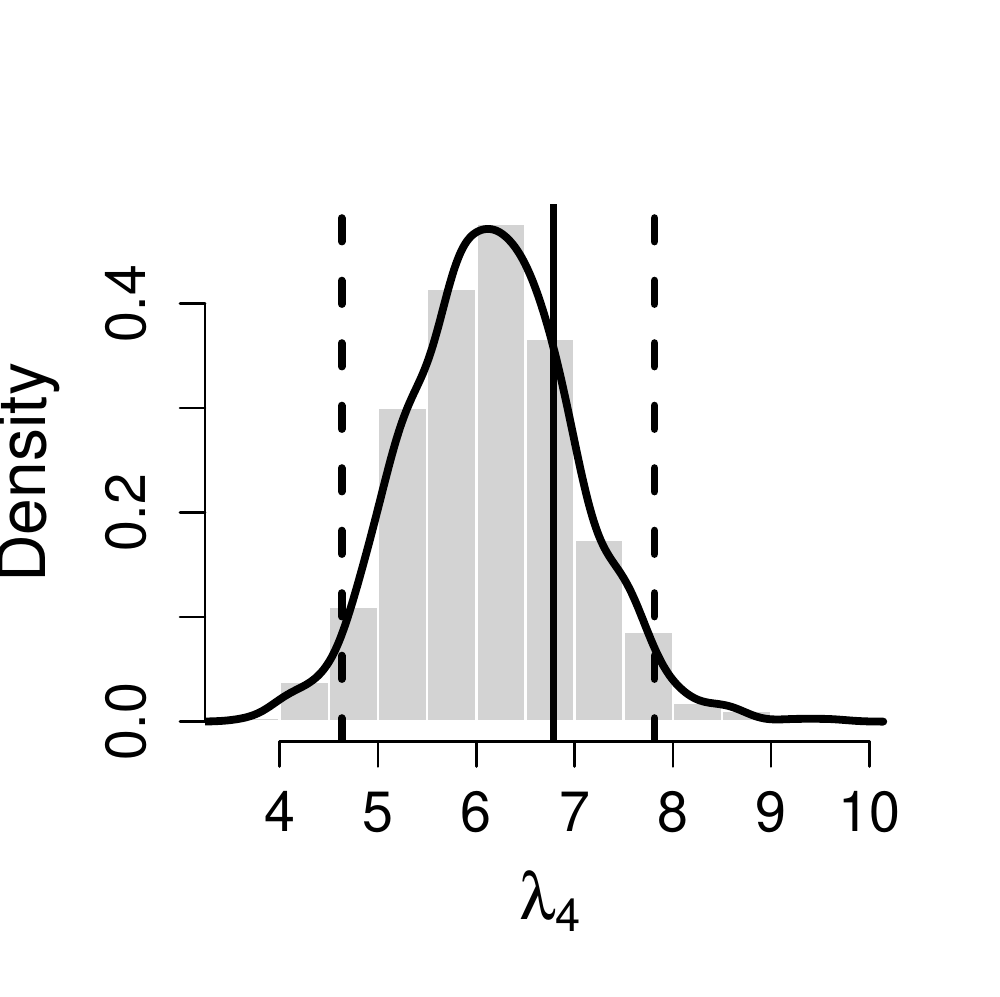}}\\
\hspace{-0.5 cm}{\includegraphics[scale=0.4]{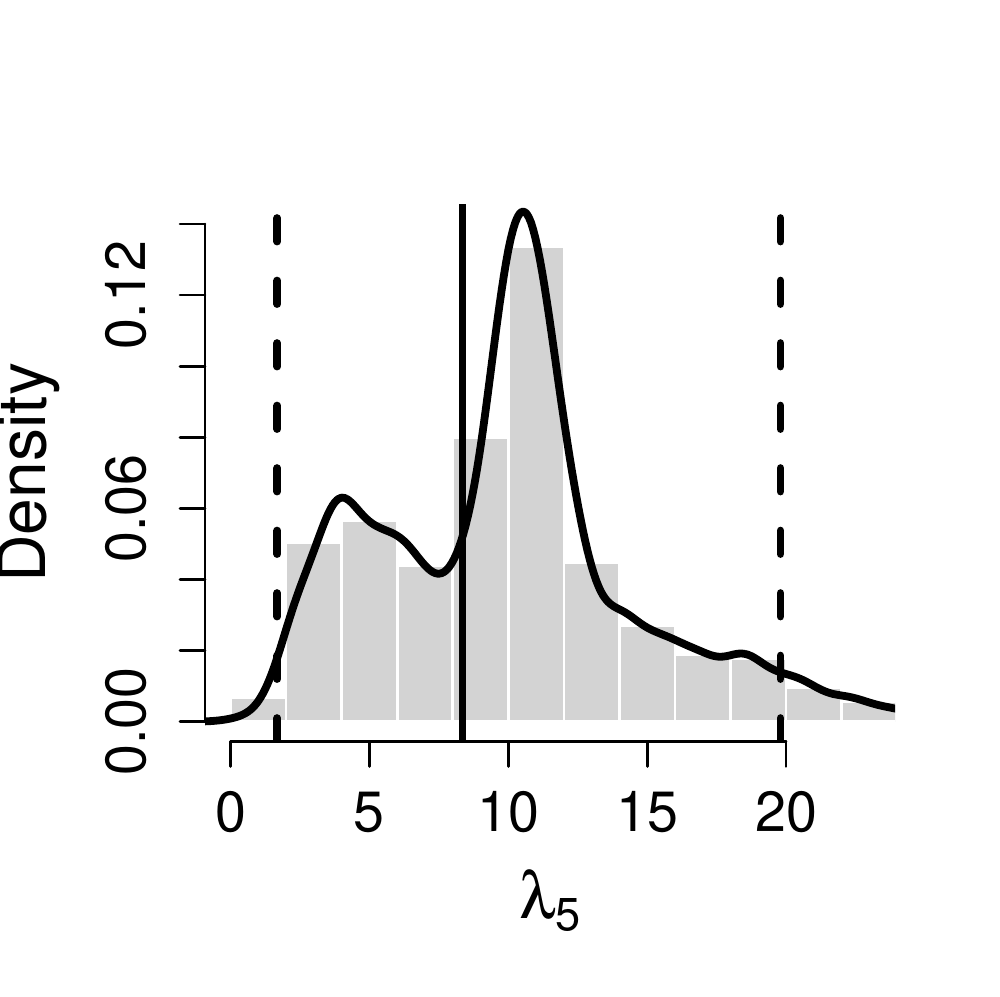}}
\hspace{-0.5 cm}{\includegraphics[scale=0.4]{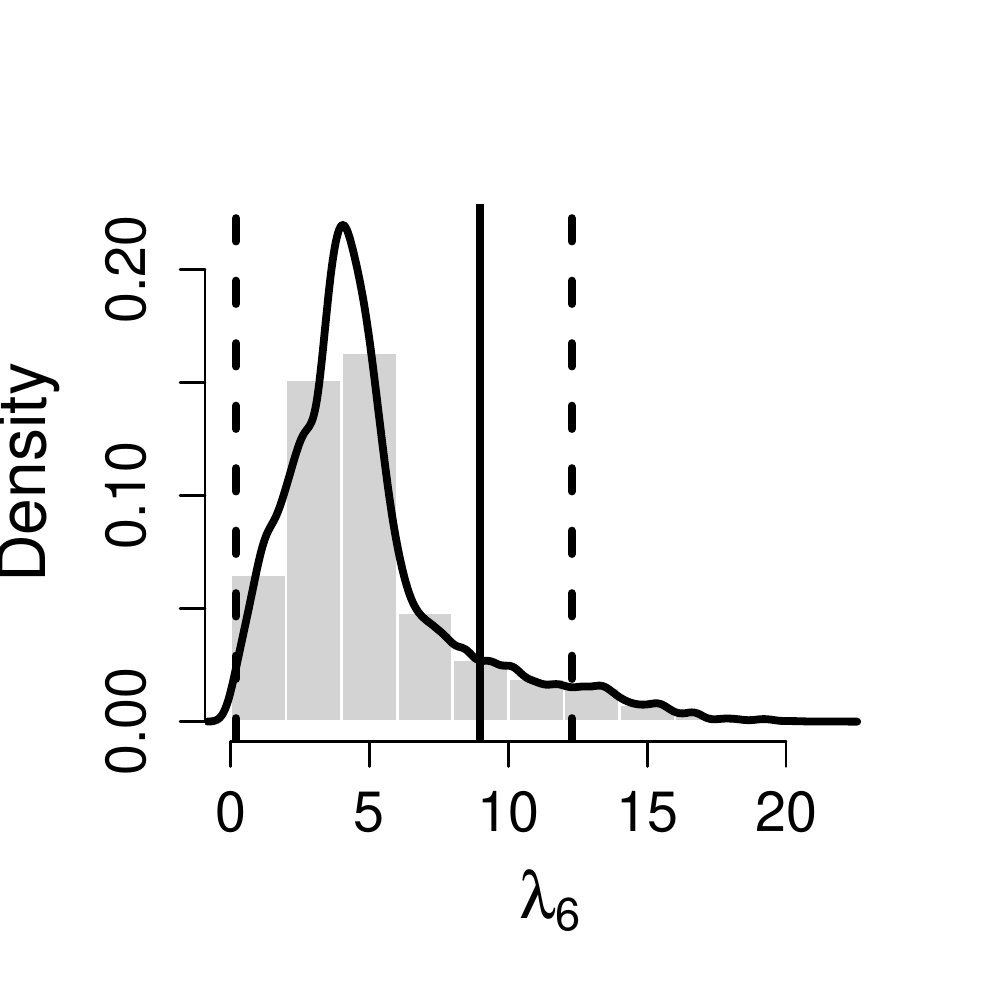}}
\hspace{-0.5 cm}{\includegraphics[scale=0.4]{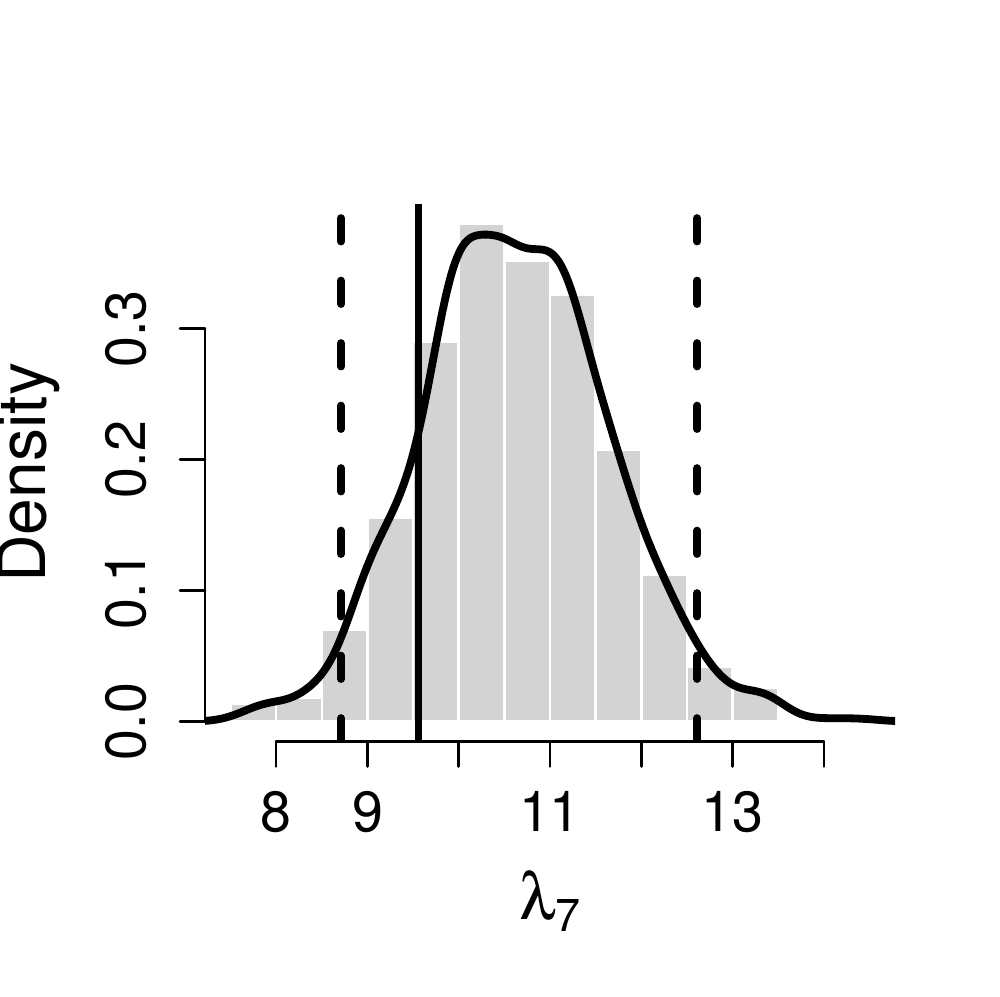}}
\hspace{-0.5 cm}{\includegraphics[scale=0.4]{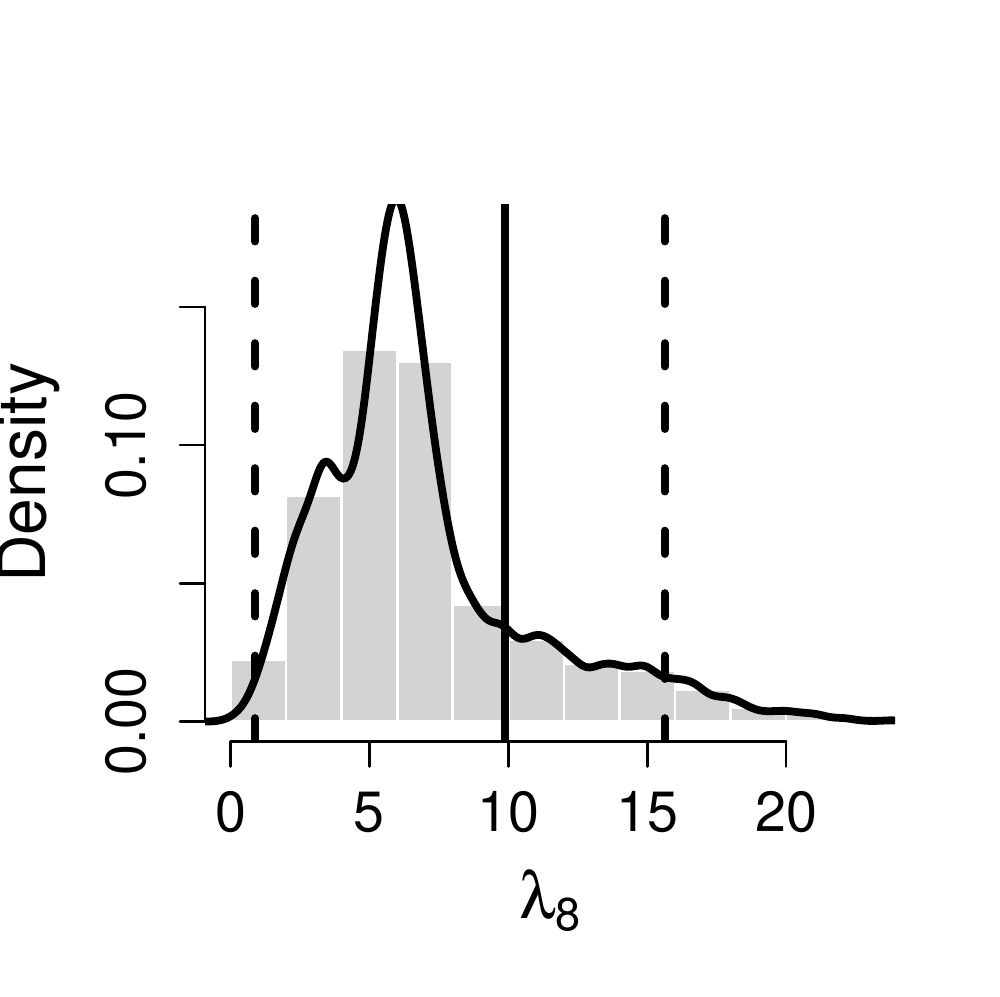}}
\end{tabular}\vspace{-0.3 cm}\caption{\it Posterior densities for certain model parameters and the population total $T$ for an artificial population. The vertical solid line is the true value fixed in the simulation, and the dashed line is the 95\% HPD interval.} \label{post_1}
\end{figure}

 For the RJMCMC simulations, we generated 100,000 samples from the posterior distribution, discarded the first 10,000, and then thinned the chain by taking every 90th sample value. Figure \ref{post_1} displays the histogram with the posterior densities of $\alpha$, $\beta$, $\nu$, $\bflambda$ and $T$ for the generated population. The posterior densities of $\bflambda$'s components are conditional on the posterior samples, whose estimated value of  $R$ is equal to eight. The solid and the dashed lines represent the true value and the 95\% highest posterior density (HPD) interval, respectively. It should be noted that most of the parameters are well estimated, with their true value within the 95\% HPD interval.

It should be noted that some $\lambda_j$'s associated with unobserved networks have bimodal posterior distributions and lower precision.  This behavior is something expected in the posterior densities of mixture model parameters obtained by RJMCMC and is generally associated with the labeling at each sweep-see \cite{richardson1997bayesian}. For instance, let us consider the case of two normal distributions, unambiguously labeled. The posterior distribution of the two means could overlap, but the extent of the overlap depends on its separation and the sample size. When the means are well separated, labels of the realizations from the posterior by ordering their means generally coincide with the population ones. As the separation reduces, ``label switching" may occur. This problem can be minimized by choosing to order other parameters of the mixture components, for example, the variance. In our case, this bimodality does not appear in all the simulations, only on ones generated by the $\lambda_j$'s that are not well separated. Nevertheless, the bimodality influences neither the convergence of the other parameters nor the most important quantity: the total $T$.

The $\lambda_j$'s associated with the sampled networks present better estimates than the  $\lambda_j$'s associated with the non-sampled networks. This result is expected because we have specific information for the sampled networks.

Two other diagnostics were used to show that the convergence was achieved: the Geweke and the Raftery-Lewis. The first was proposed by \cite{geweke1992} and is based on a test for equality of the means of the first and last part of the Markov chain. If the samples are drawn from the stationary distribution of the chain, the two means are equal, and Geweke's statistic has an asymptotically standard normal distribution. The second was proposed in \cite{raftery1992practical} and calculates the number of iterations required to estimate a quantile with a desired accuracy and with a certain probability. The minimum length is the required sample size for a chain with no correlation between consecutive samples. An estimate dependence factor of the extent to which autocorrelation inflates the required sample size is also provided. Values for the factor that are larger than $5$ indicate strong autocorrelation, which may be due to a poor choice of starting value, high posterior correlations or stickiness of the MCMC algorithm. Table \ref{criteria_conv} presents the value of Geweke's statistic and the value of the dependence factor. The results for both criteria indicate that the MCMC chains have converged.
{\footnotesize\begin{table*}[h]\caption{ {\it Geweke and Raftery-Lewis convergence diagnostics for all of the parameters estimated for the artificial population.}}
\begin{center}
\begin{tabular}{c|cccccccccccccc}
& $\alpha$ & $\beta$ & $\nu$ & $T$ & $\lambda_1$ & $\lambda_2$ & $\lambda_3$ & $\lambda_4$ & $\lambda_5$ & $\lambda_6$ & $\lambda_7$ & $\lambda_8$\\\hline
Geweke & 0.7 & -0.4 & -1.6 & 0.4 & 1.4 & -1.3 & 1.4 &-0.4 & 1.5 & 1.5 & 1.2 & 1.5\\
R-L & 1.3 & 1.1 & 1.1 & 1.8 & 0.9 & 1.0 & 1.0 & 1.0 & 0.9 & 1.0 & 1.1 & 1.1
	\end{tabular}\label{criteria_conv}
		\end{center}
\end{table*}}

\section{Simulation study}\label{simul_study}

To examine the performance of the Bayesian estimator and the influence of the different prior models on the Poisson parameters, we sampled several simulated clustered populations and obtained samples from the posterior distributions of the model parameters and population parameters. The population estimates were then compared with the true values to evaluate the model's performance.

\subsection{ Simulation scenarios}\label{dif_set}

 We generated $500$ populations for each scenario that we considered. Twelve scenarios were created by varying the values of $N$, $R$ and $X$, as well as varying the $\bflambda$ components. The values of parameters $(\alpha,\beta)$ were fixed such that their combinations expressed different degrees of rare and clustered populations. For the first simulation study, we considered only the independent prior for $\bflambda$; thus, we generated the values of the components of $\bflambda$ as a Gamma distribution with  $d=1.1$ and $\nu=0.13$.  These values of $d$ and $\nu$ ensure that the generated populations provide heterogeneous networks. Finally, an adaptive cluster sample was selected from each population, with the first stage as a $5\%$  simple random sample without replacement.

Table \ref{table_mixtmod} shows summary statistics with some frequentist measures of the posterior distributions of the model parameters after reaching convergence  for each of the twelve evaluated scenarios. It reports the relative mean square error (RMSE), the relative absolute error (RAE), the empirical nominal coverage of the 95\% HPD intervals measured in percentages and the respective widths averaged over the $500$ simulations. In particular, to facilitate future comparisons, the widths presented for the total $T$ and for $\bflambda_s$ and $\bflambda_{\bar{s}}$ are expressed in ratio form relative to their true values. The results for $\lambda_j$'s are separately summarized for $\bflambda_s$ and $\bflambda_{\bar{s}}$.

\begin{table*}[h]\caption{ {\it Summary measurements for the point and interval estimates of the model and population parameters over 500 simulations for different values of $\alpha$, $\beta$ and $N$.}}\vspace{0 cm}
\hspace{-1 cm}\begin{center}
{\footnotesize\begin{tabular}{c|cccccc|cccccccc} \hline
&\multicolumn{12}{c}{$N=200$}\\\hline
&\multicolumn{6}{c|}{$(\alpha,\beta)=(0.10,0.10)$}&\multicolumn{6}{|c}{$(\alpha,\beta)=(0.10,0.15)$}\\\hline
& $T$ & $\alpha$ & $\beta$ & $\nu$ & $\bflambda_s$ & $\bflambda_{\bar{s}}$ & $T$ & $\alpha$ & $\beta$ & $\nu$ & $\bflambda_s$ & $\bflambda_{\bar{s}}$\\\hline
RMSE & 0.21 & 0.38 & 0.53 & 0.56 & 0.03 & 0.29 & 0.22 & 0.29 & 0.29 & 0.39 & 0.03 & 0.28\\
RAE  & 0.35 & 0.17 & 0.25 & 0.60 & 0.12 & 0.46 & 0.36 & 0.16 & 0.35 & 0.47 & 0.13 & 0.45\\
Cov. & 95.0 & 91.1 & 96.7 & 89.5 & 91.7 & 87.8 & 93.8 & 93.7 & 98.1 & 89.7 & 90.3 & 87.7\\
Wid. & 1.60 & 0.20 & 0.31 & 0.28 & 0.58 & 1.23 & 1.60 & 0.19 & 0.31 & 0.28 & 0.57 & 1.26 \\\hline
&\multicolumn{6}{c|}{$(\alpha,\beta)=(0.15,0.1)$}&\multicolumn{6}{|c}{$(\alpha,\beta)=(0.15,0.15)$}\\\hline
RMSE & 0.09 & 0.20 & 0.50 & 0.22 & 0.02 & 0.31 & 0.06 & 0.10 & 0.19 & 0.32 & 0.02 & 0.27\\
RAE & 0.24 & 0.31 & 0.45 & 0.40 & 0.11 & 0.46 & 0.21 & 0.27 & 0.21 & 0.47 & 0.10 & 0.41\\
Cov. & 94.6 & 90.9 & 97.1 & 90.2 & 93.6 & 89.1 & 97.3 & 97.0 & 98.5 & 90.5 & 94.1 & 89.8 \\
Wid. & 1.22 & 0.19 & 0.21 & 0.22 & 0.50 & 1.33 & 1.24 & 0.20 & 0.23 & 0.21 & 0.56 & 1.51\\\hline
&\multicolumn{12}{c}{$N=400$}\\\hline
&\multicolumn{6}{c|}{$(\alpha,\beta)=(0.10,0.10)$}&\multicolumn{6}{|c}{$(\alpha,\beta)=(0.10,0.15)$}\\\hline
& $T$ & $\alpha$ & $\beta$ & $\nu$ & $\bflambda_s$ & $\bflambda_{\bar{s}}$ & $T$ & $\alpha$ & $\beta$ & $\nu$ & $\bflambda_s$ & $\bflambda_{\bar{s}}$\\\hline
RMSE & 0.06 & 0.15 & 0.42 & 0.14 & 0.02 & 0.29 & 0.05 & 0.08 & 0.15 & 0.10 & 0.02 & 0.31\\
RAE  & 0.21 & 0.32 & 0.35 & 0.28 & 0.10 & 0.43 &0.20 & 0.23 & 0.29 & 0.21 & 0.12 & 0.43\\
Cov. & 96.7 & 91.1 & 96.0 & 90.8 & 94.2 & 91.0 & 96.8 & 95.1 & 98.1 & 90.5 & 94.3 & 91.8\\
Wid. & 1.04 & 0.09 & 0.20 & 0.19 & 0.47 & 1.38 & 1.05 & 0.10 & 0.21 & 0.18 & 0.55 & 1.64 \\\hline
&\multicolumn{6}{c|}{$(\alpha,\beta)=(0.15,0.1)$}&\multicolumn{6}{|c}{$(\alpha,\beta)=(0.15,0.15)$}\\\hline
RMSE & 0.04 & 0.06 & 0.35 & 0.04 & 0.02 & 0.30 & 0.05 & 0.03 & 0.15 & 0.03 & 0.02 & 0.36\\
RAE & 0.18 & 0.18 & 0.39 & 0.18 & 0.09 & 0.42 & 0.20 & 0.15 & 0.21 & 0.15 & 0.10 & 0.43\\
Cov. & 93.4 & 91.2 & 96.9 & 96.7 & 94.2 & 93.9 & 92.4 & 97.0 & 98.7 & 96.5 & 93.5 & 95.6 \\
Wid. & 0.79 & 0.11 & 0.15 & 0.14 & 0.45 & 1.43 & 0.77 & 0.11 & 0.16 & 0.13 & 0.51 & 1.77\\\hline
&\multicolumn{12}{c}{$N=600$}\\\hline
&\multicolumn{6}{c|}{$(\alpha,\beta)=(0.10,0.10)$}&\multicolumn{6}{|c}{$(\alpha,\beta)=(0.10,0.15)$}\\\hline
& $T$ & $\alpha$ & $\beta$ & $\nu$ & $\bflambda_s$ & $\bflambda_{\bar{s}}$ & $T$ & $\alpha$ & $\beta$ & $\nu$ & $\bflambda_s$ & $\bflambda_{\bar{s}}$\\\hline
RMSE & 0.04 & 0.05 & 0.25 & 0.10 & 0.02 & 0.32 & 0.05 & 0.03 & 0.11 & 0.09 & 0.02 & 0.35\\
RAE  & 0.17 & 0.17 & 0.28 & 0.12 & 0.09 & 0.42 & 0.20 & 0.14 & 0.26 & 0.11 & 0.11 & 0.42\\
Cov. & 96.3 & 91.8 & 98.1 & 98.0 & 93.5 & 93.1 & 92.8 & 97.5 & 98.3 & 97.0 & 93.8 & 96.1\\
Wid. & 0.79 & 0.08 & 0.22 & 0.20 & 0.46 & 1.40 & 0.78 & 0.08 & 0.23 & 0.19 & 0.52 & 1.70 \\\hline
&\multicolumn{6}{c|}{$(\alpha,\beta)=(0.15,0.10)$}&\multicolumn{6}{|c}{$(\alpha,\beta)=(0.15,0.15)$}\\\hline
RMSE & 0.05 & 0.04 & 0.21 & 0.06 & 0.01 & 0.37 & 0.09 & 0.08 & 0.06 & 0.05 & 0.02 & 0.35\\
RAE & 0.19 & 0.17 & 0.30 & 0.09 & 0.09 & 0.44 & 0.29 & 0.24 & 0.18 & 0.09 & 0.10 & 0.43\\
Cov. & 90.4 & 91.1 & 98.7 & 98.9 & 95.3 & 96.0 & 90.0 & 90.5 & 98.8 & 98.4 & 95.5 & 96.8 \\
Wid. & 0.78 & 0.08 & 0.17 & 0.18 & 0.43 & 1.49 & 0.53 & 0.08 & 0.20 & 0.17 & 0.53 & 1.79\\\hline
\end{tabular}}\label{table_mixtmod}
\end{center}
\end{table*}

In general, the parameters are well estimated. The coverage of the 95\% HPD intervals is close to the nominal level. The RMSE and RAE are small for all the parameters, except for $\beta$ in certain specific cases. However, there is no significant impact on the prediction of the total $T$, which is our main interest. As expected, the results for $\lambda_j$ obtained with the samples containing the network $j$ show smaller errors and are more precise than the results that consider the samples in which the network $j$ was not observed. As the value of $N$ increases, the RMSEs and RAEs of most of the parameters decrease. This phenomenon may occur because the number of non-empty networks increases with $N$, improving the estimates of $\alpha$ and $\beta$ and consequently of the other parameters.  However, for the same reason, for a fixed value of $N$, the errors decrease as the values of $\alpha$ and $\beta$ increase.

\begin{figure}[h!]
\begin{tabular}{c}
\hspace{-0.5 cm}\subfigure[RE - $\bflambda_s$]{\includegraphics[scale=0.32]{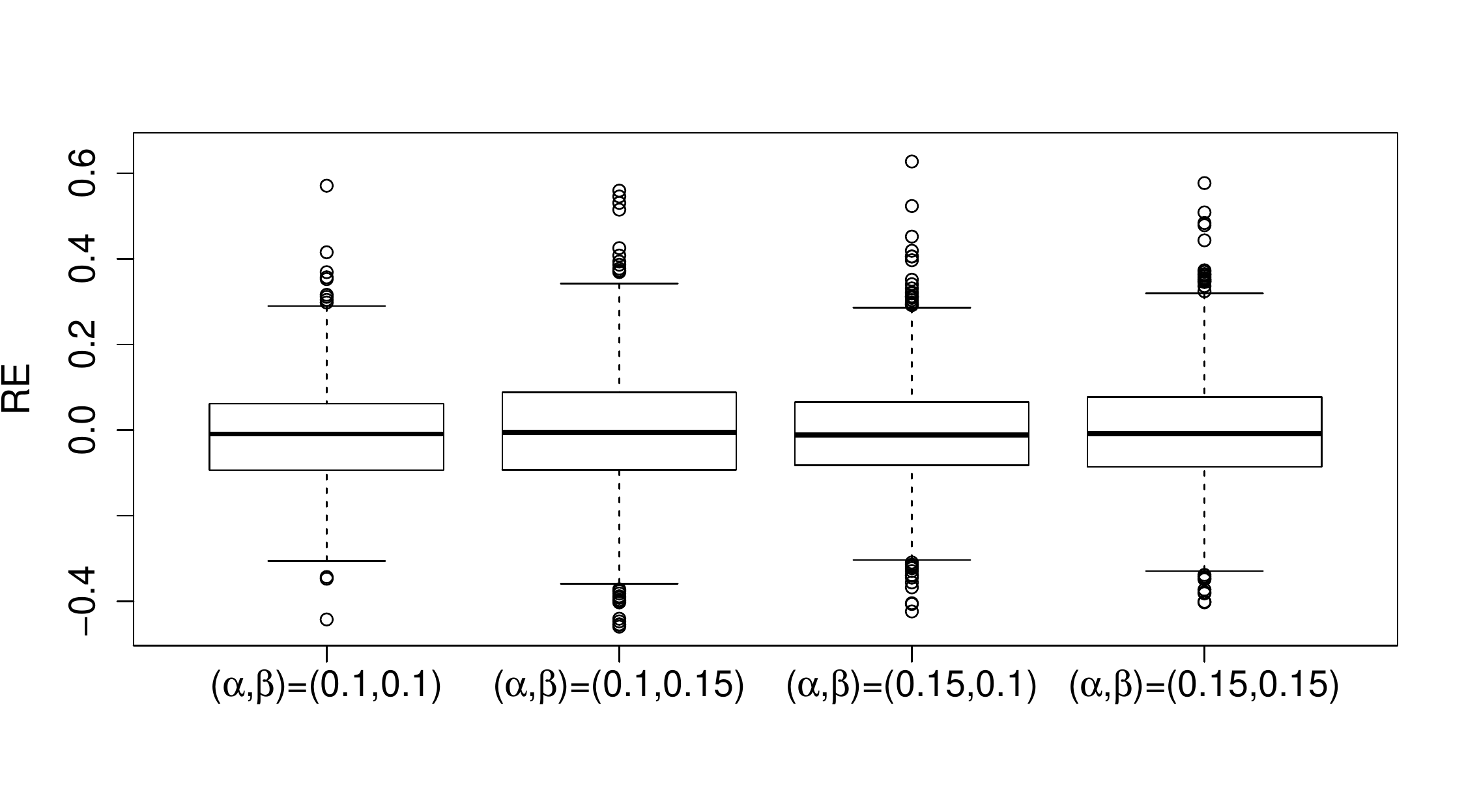}}\hspace{-0.3 cm}
\subfigure[RE - $\bflambda_{\bar{s}}$]{\includegraphics[scale=0.32]{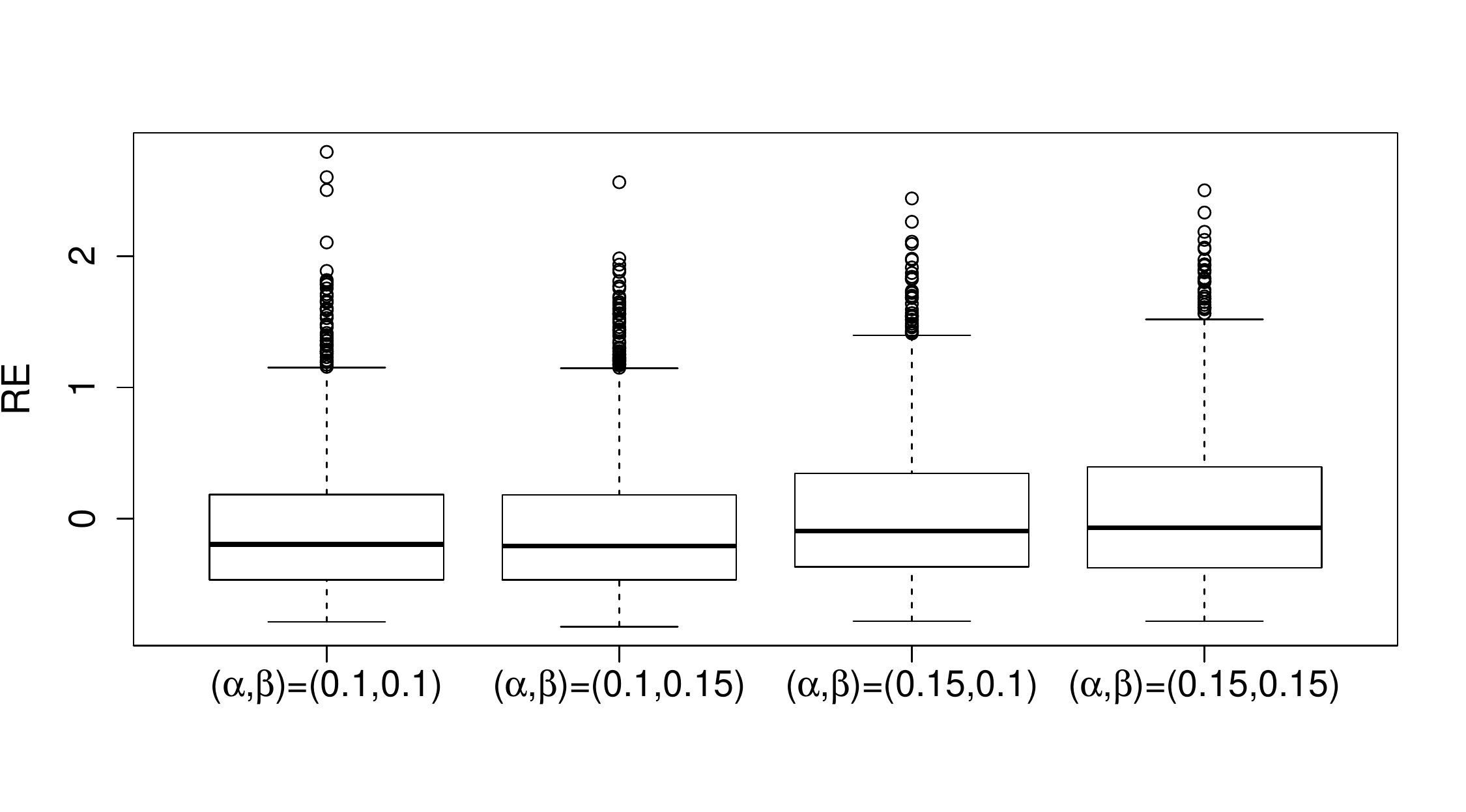}}
\end{tabular}
\vspace{-0.3 cm}\caption{{\it Relative errors for $\bflambda_s$ and $\bflambda_{\bar{s}}$ over 500 simulations, for $N=400$ and different values of $\alpha$ and $\beta$.}}\label{summary_lambda_together}
\end{figure}

It is not possible to present the frequentist properties for each $\lambda_j$ because the value of $R$ was not fixed over the simulations. Figure  \ref{summary_lambda_together} presents the relative errors (REs) of $\bflambda_s$ and $\bflambda_{\bar{s}}$ for all the networks and all the simulations, for different values of $\alpha$ and $\beta$ and for $N=400$. Note that, in all cases, the RE is approximately zero and is smaller for $\bflambda_s$, as expected. Note also that $\bflambda_{\bar{s}}$ is slightly underestimated.

The $500$ populations were previously generated by fixing the parameters of $\lambda_j$'s Gamma distribution at $d=1.1$ and $\nu=0.13$, yielding a mean of $8.5$ and a CV of 95\%. The aim here is to evaluate the performance of the proposed model with respect to the level of homogeneity.  We consider two extra values of CVs: 25\% and 50\%, with the means fixed at $8.5$ for both. Then, we calculate the two sets of values of $d$ and $\nu$.  When the CV is fixed at $50\%$, we obtain $d=4$ and $\nu=0.47$; when the CV equals $25\%$, the result is $d=16$ and $\nu=1.89$.

\begin{figure}[h!]
\begin{center}
\begin{tabular}{c}
{\includegraphics[scale=0.4]{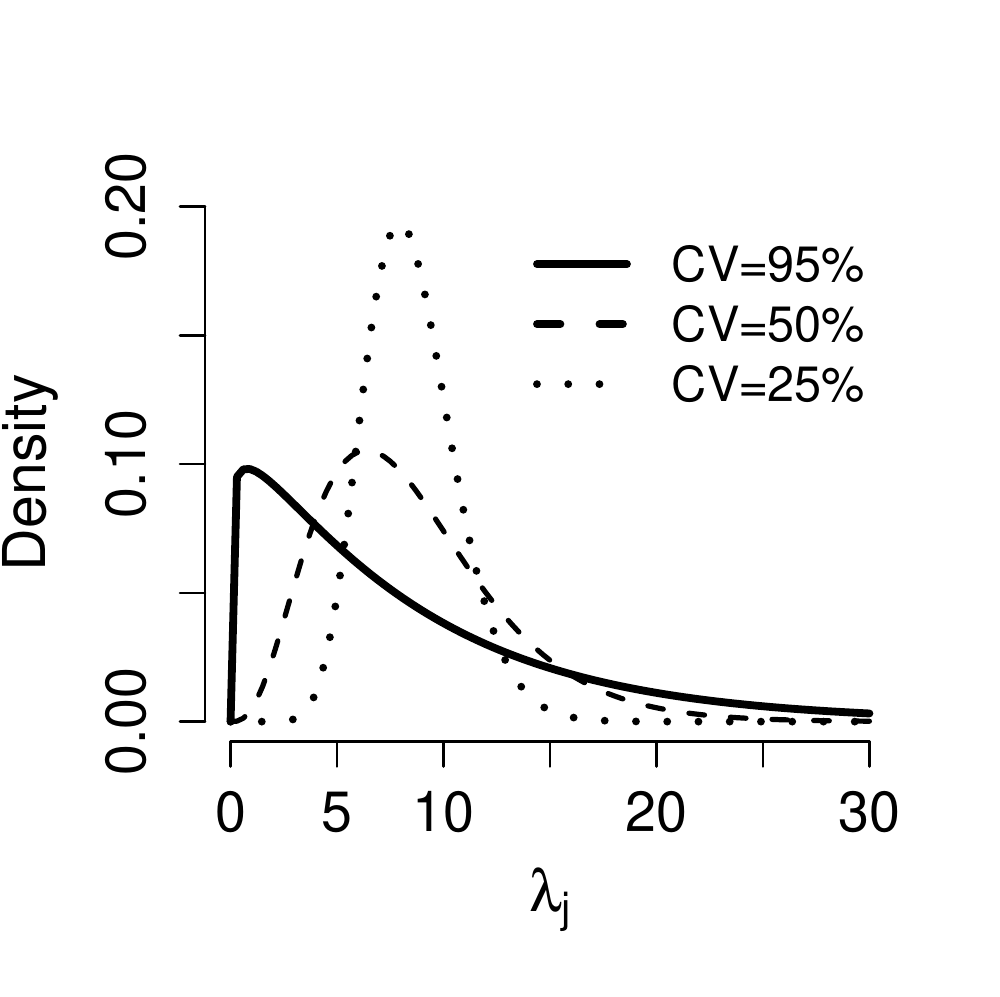}}
\end{tabular}
\end{center}
\vspace{-0.8 cm}\caption{{\it Prior distributions to $\lambda_j$ used in the simulations, varying the value of the CV of the distribution.}}\label{gammas}
\end{figure}

Figure \ref{gammas} displays the densities of $\lambda_j$ for each fixed value of the CV. Note that, as the CV decreases, the prior distribution for $\lambda_j$ becomes more concentrated and symmetrical around the mean of the distribution; consequently, the networks will become more homogeneous with respect to the total in their units.
 
We generated two other sets of populations, fixing the CVs of the $\lambda_j$ distributions to $50\%$ and $25\%$, respectively. The population size was set at $N=400$, and a $5\%$ adaptive sample was taken from it.

Table \ref{table_mixtmod_CV} presents summary measurements of the estimators over the 500 populations generated for the two values considered for the CV. It should be noted that, even for the more homogeneous cases, the proposed model (\ref{model_mixt_prop}) has a good performance, resulting in parameter estimates with small errors and $95\%$ HPD intervals with coverage probability near the fixed nominal level.

\begin{table*}[h]\caption{ {\it Summary measurements for the point and interval estimation of the model parameters over $500$ simulations, varying the level of homogeneity in $\bflambda$, expressed as the coefficient of variation of its distribution, for $N=400$.}}\vspace{0 cm}
\hspace{-1 cm}\begin{center}
{\footnotesize\begin{tabular}{c|cccccc|cccccccc} \hline
&\multicolumn{12}{c}{$CV = 50\%$}\\\hline
&\multicolumn{6}{c|}{$(\alpha,\beta)=(0.10,0.10)$}&\multicolumn{6}{|c}{$(\alpha,\beta)=(0.10,0.15)$}\\\hline
& $T$ & $\alpha$ & $\beta$ & $\nu$ & $\bflambda_s$ & $\bflambda_{\bar{s}}$ & $T$ & $\alpha$ & $\beta$ & $\nu$ & $\bflambda_s$ & $\bflambda_{\bar{s}}$\\\hline
RMSE & 0.13 & 0.15 & 0.52 & 0.16 & 0.02 & 0.04 & 0.06 & 0.09 & 0.18 & 0.10 & 0.02 & 0.03\\
RAE  & 0.26 & 0.32 & 0.27 & 0.30 & 0.10 & 0.15 & 0.18 & 0.24 & 0.36 & 0.23 & 0.11 & 0.15\\
Cov. & 95.3 & 87.2 & 97.0 & 95.3 & 94.7 & 97.0 & 96.7 & 95.0 & 98.2 & 95.0 & 94.5 & 97.6\\
Wid. & 1.38 & 0.11 & 0.26 & 0.91 & 0.51 & 1.27 & 1.24 & 0.11 & 0.27 & 0.82 & 0.55 & 1.31 \\\hline
&\multicolumn{6}{c|}{$(\alpha,\beta)=(0.15,0.1)$}&\multicolumn{6}{|c}{$(\alpha,\beta)=(0.15,0.15)$}\\\hline
RMSE & 0.03 & 0.04 & 0.40 & 0.08 & 0.02 & 0.03 & 0.03 & 0.03 & 0.10 & 0.06 & 0.02 & 0.03\\
RAE  & 0.15 & 0.15 & 0.50 & 0.21 & 0.10 & 0.12 & 0.16 & 0.14 & 0.26 & 0.18 & 0.10 & 0.13\\
Cov. & 96.5 & 94.7 & 97.3 & 97.8 & 95.6 & 98.0 & 95.8 & 97.3 & 98.0 & 97.5 & 95.8 & 97.9 \\
Wid. & 0.95 & 0.11 & 0.23 & 0.75 & 0.48 & 1.28 & 0.92 & 0.11 & 0.24 & 0.70 & 0.53 & 1.36\\\hline
&\multicolumn{12}{c}{$CV = 25\%$}\\\hline
&\multicolumn{6}{c|}{$(\alpha,\beta)=(0.10,0.10)$}&\multicolumn{6}{|c}{$(\alpha,\beta)=(0.10,0.15)$}\\\hline
RMSE & 0.09 & 0.30 & 0.50 & 0.36 & 0.03 & 0.08 & 0.05 & 0.18 & 0.12 & 0.34 & 0.03 & 0.08\\
RAE  & 0.23 & 0.48 & 0.37 & 0.47 & 0.13 & 0.24 & 0.19 & 0.37 & 0.29 & 0.44 & 0.14 & 0.26\\
Cov. & 89.7 & 86.8 & 98.0 & 75.0 & 85.7 & 82.2 & 94.7 & 90.1 & 98.2 & 74.9 & 85.7 & 81.0\\
Wid. & 0.96 & 0.12 & 0.25 & 3.01 & 0.47 & 0.70 & 0.91 & 0.12 & 0.27 & 2.83 & 0.51 & 0.75 \\\hline
&\multicolumn{6}{c|}{$(\alpha,\beta)=(0.15,0.1)$}&\multicolumn{6}{|c}{$(\alpha,\beta)=(0.15,0.15)$}\\\hline
RMSE & 0.03 & 0.08 & 0.41 & 0.25 & 0.02 & 0.03 & 0.04 & 0.05 & 0.07 & 0.19 & 0.02 & 0.04\\
RAE  & 0.14 & 0.22 & 0.49 & 0.34 & 0.10 & 0.15 & 0.17 & 0.15 & 0.21 & 0.24 & 0.11 & 0.17\\
Cov. & 96.6 & 91.7 & 97.5 & 80.8 & 94.6 & 94.4 & 91.9 & 92.5 & 98.3 & 83.2 & 93.3 & 94.8 \\
Wid. & 0.70 & 0.12 & 0.22 & 2.48 & 0.46 & 0.74 & 0.70 & 0.12 & 0.23 & 2.25 & 0.50 & 0.79\\\hline
		\end{tabular}}\label{table_mixtmod_CV}
		\end{center}
\end{table*}

\clearpage

In particular, the relative errors of $T$ do not vary much with the values of the CV, except when $(\alpha,\beta)=(0.10,0.10)$, for which, on average, smaller numbers of non-empty networks in the generated populations are found. In addition, the relative errors for $\bflambda_{\bar{s}}$ are smaller than the errors obtained when CV is fixed in $95\%$, though the errors for $\nu$ become larger. Furthermore, as the CV decreases, the empirical coverage of nominal 95\% HPD intervals is underestimated, mainly with respect to $\nu$ and $\bflambda$.

\subsection{Prior sensitivity analysis}

In this section, we compare the performance of the two prior distributions considered for $\bflambda$.  To obtain simulation results for each component $\lambda_j$ of $\bflambda$ using a different method from the previous section, the values of $R$ were fixed. The population size was set at $N=400$, and $(\alpha,\beta)=(0.15,0.10)$. These settings were chosen to provide rare and clustered populations as much as possible. Then, we conducted a large number of simulations until we reached $500$ populations with $R=5$; another 500 populations were generated with $R=6$, followed by another 500 populations with $R=7$. We consider only these values of $R$ because the others have much lower probabilities of being generated in this simulation scenario with $(\alpha,\beta)=(0.15,0.10)$. Furthermore, because we were specifying two different priors for $\bflambda$, we fixed the $\bflambda$'s components at $(4.5,4.8,8.0,11.3,13.8)$ for $R=5$, at $(3.9,6.4,6.9,7.1,10.5,14.8)$ for $R=6$ and at $(4.8,7.4,9.5,10.1,11.4,11.7,14.5)$ for $R=7$. These values were generated from a uniform distribution defined in the interval $(3,15)$.

All results shown hereafter correspond to $100,000$ RJMCMC sweeps, after $10,000$ burn-ins; the chain was then thinned by taking every $10^{th}$ sample value. We used the same prior distribution for $\alpha$ and $\beta$ described in the previous section. For $\bflambda$, we considered the Gamma prior distribution used in the previous simulation study and the dependent prior $\lambda_j\mid \lambda_{j-1}\sim \mbox{N}_{(\lambda_{j-1},\infty)}(\lambda_{j-1},\tau)$ with $\tau\in \{1,5,10,20\}$.

\begin{figure}[h!]
\begin{center}
\begin{tabular}{c}
{\includegraphics[scale=0.4]{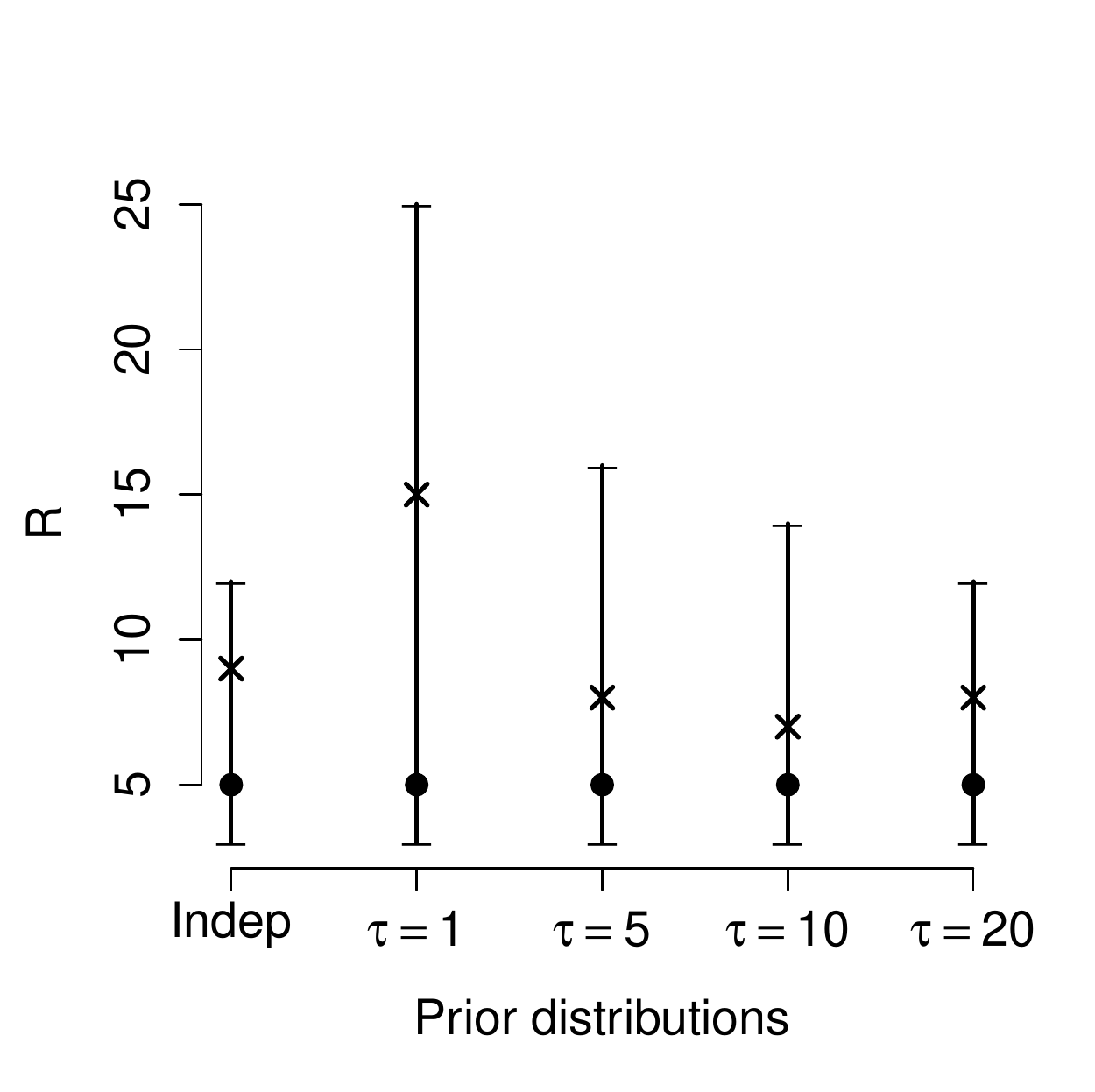}}
\end{tabular}
\end{center}
\vspace{-0.3 cm}\caption{{\it The 95\% HPD interval of $R$ for different prior distributions of $\bflambda$. The crosses represent the median of the distribution, the circle represents the true value of $R$, and the line represents the 95\% interval.}}\label{post_k}
\end{figure}

Figure \ref{post_k} shows the 95\% HPD interval obtained for $R$ for each $\bflambda$ prior assumed when we fit the model for one of the 500 populations generated. The parameter $R$ is much more sensitive to the value of $\tau$ assigned for the dependent prior.  In addition, the $R$ posterior distribution is fairly vague when $\tau=1$. However, as $\tau$ increases, this behavior is attenuated. The independent prior and the dependent one with $\tau=20$ yield approximately the same 95\% HPD interval for $R$. This behavior was observed for almost all of the 500 simulation samples. Thus, from now on, we do not consider the dependent $\bflambda$ prior with $\tau=1$.

\begin{figure}[h!]
\begin{tabular}{c}
\hspace{-0.3 cm}{\includegraphics[scale=0.46]{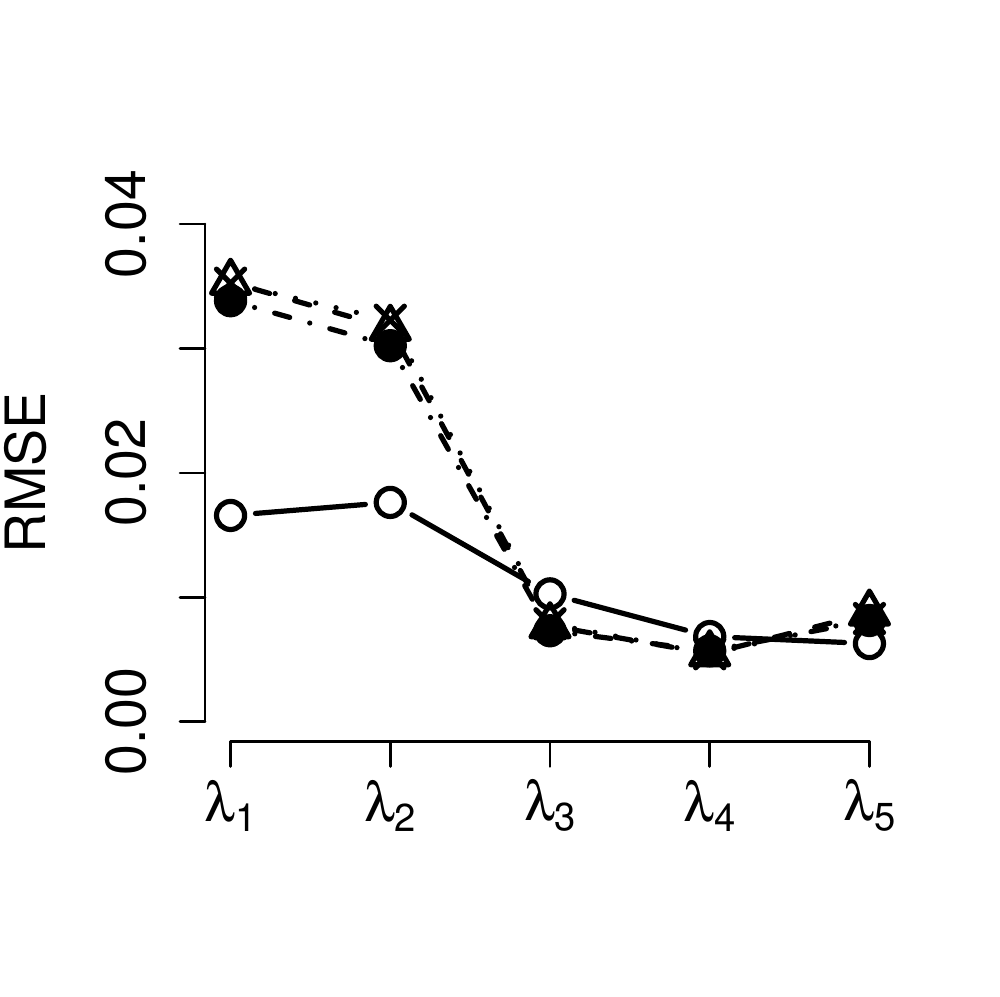}}
\hspace{-0.3 cm}\subfigure[RMSE for $\lambda_j$, $j\in s$]{\includegraphics[scale=0.46]{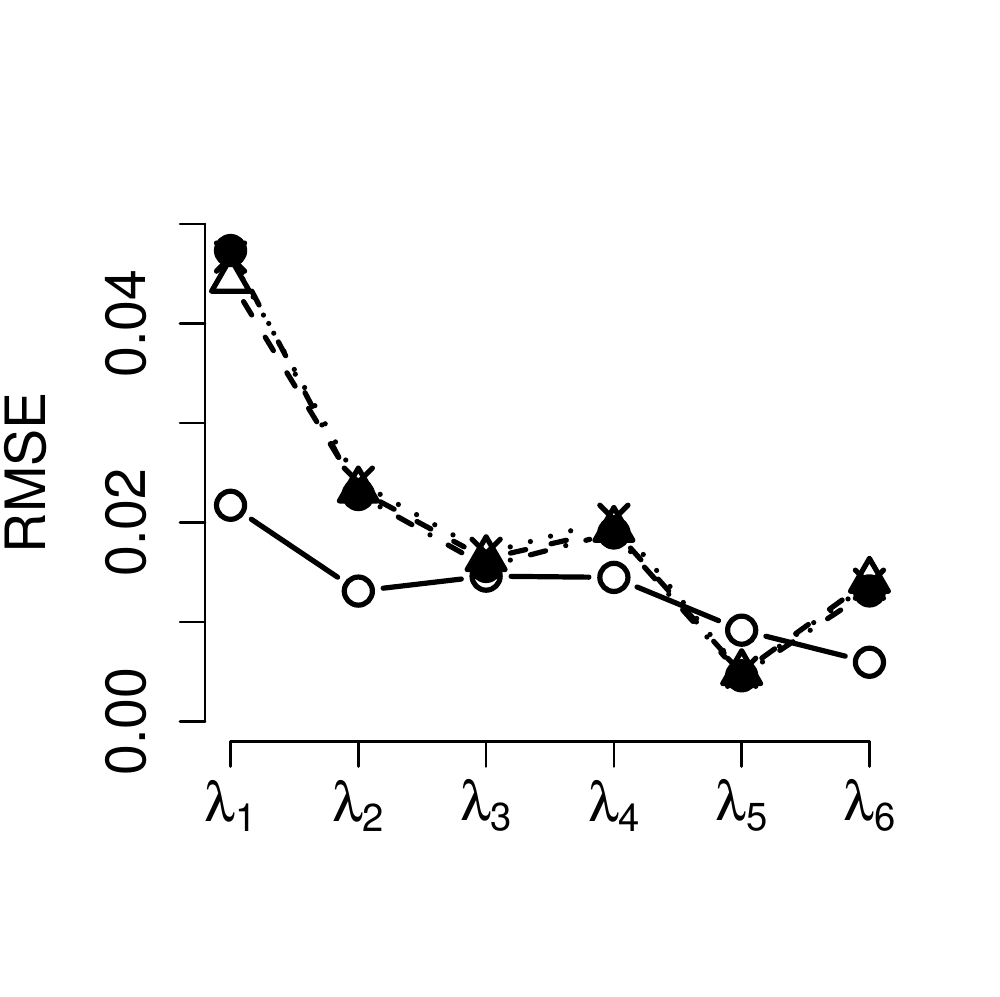}}
\hspace{-0.3 cm}{\includegraphics[scale=0.46]{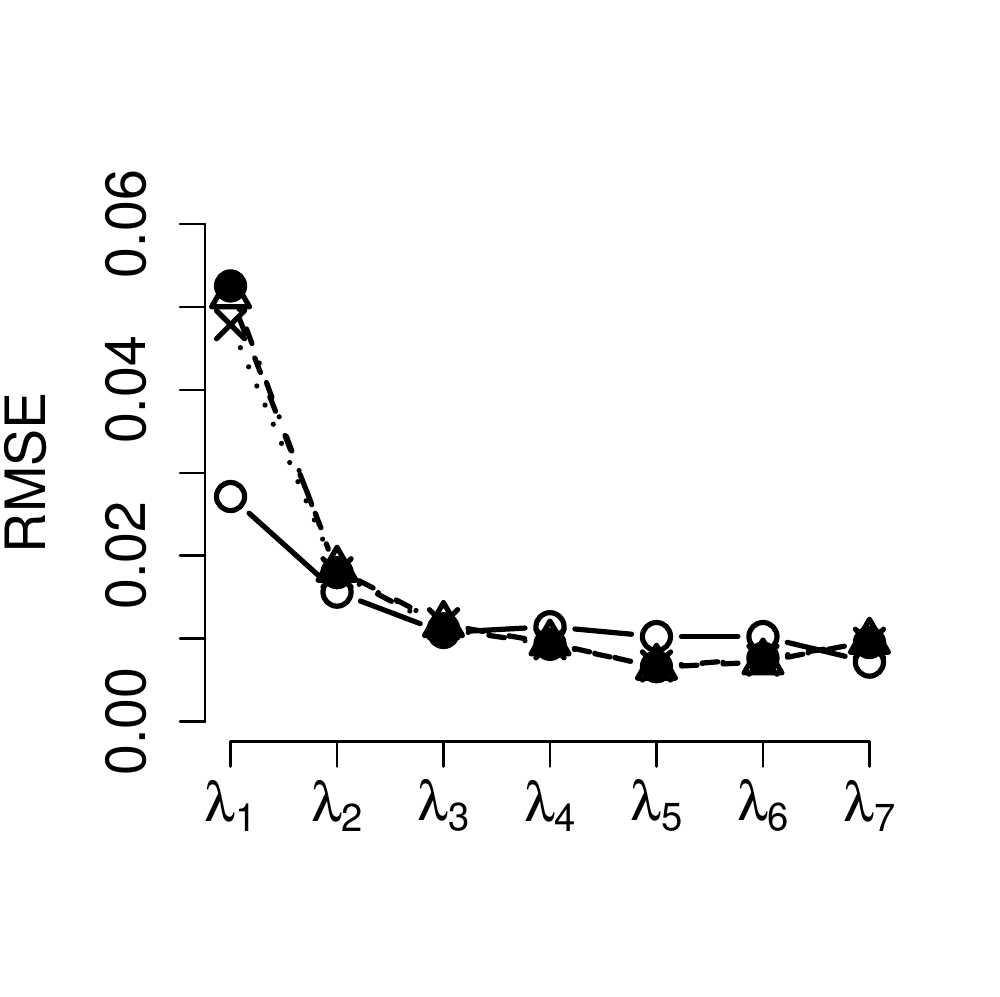}}\vspace{-1 cm}\\
\hspace{-0.3 cm}{\includegraphics[scale=0.46]{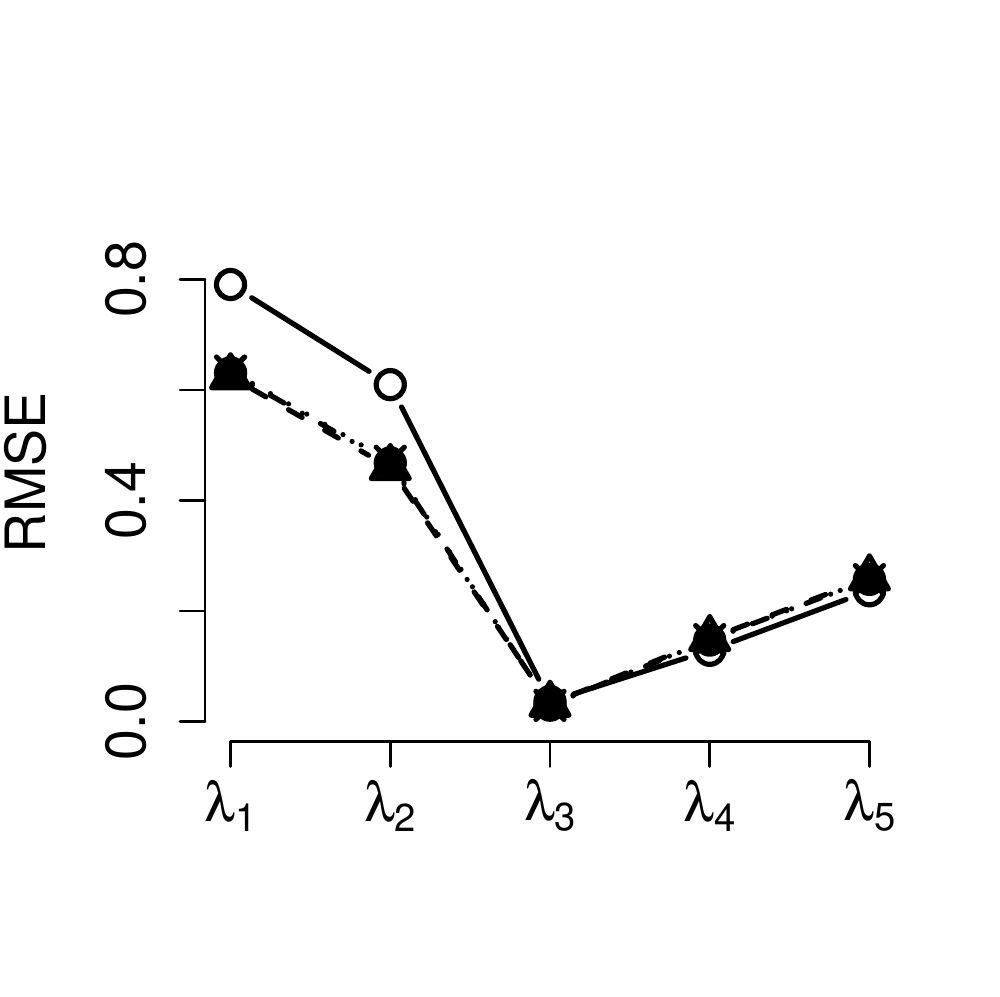}}
\hspace{-0.3 cm}\subfigure[RMSE for $\lambda_j$, $j\in \bar{s}$]{\includegraphics[scale=0.40]{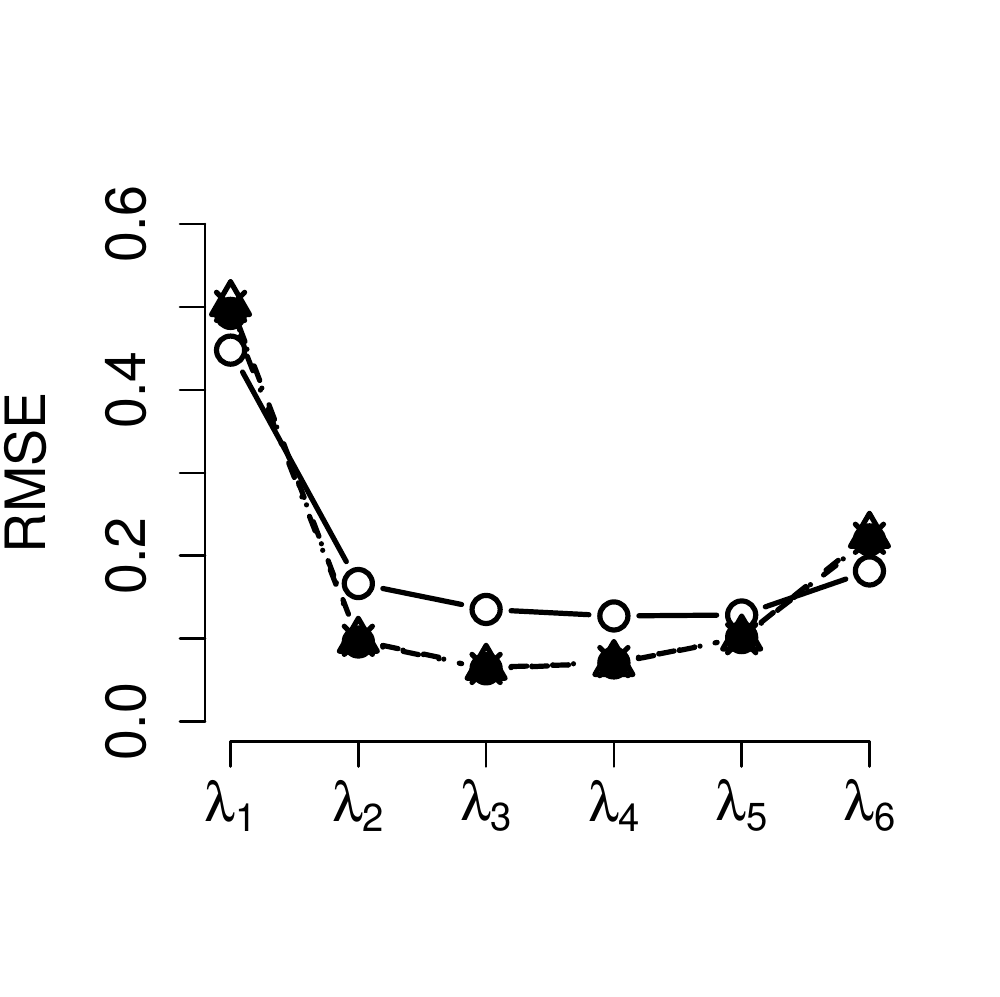}}
\hspace{-0.3 cm}{\includegraphics[scale=0.46]{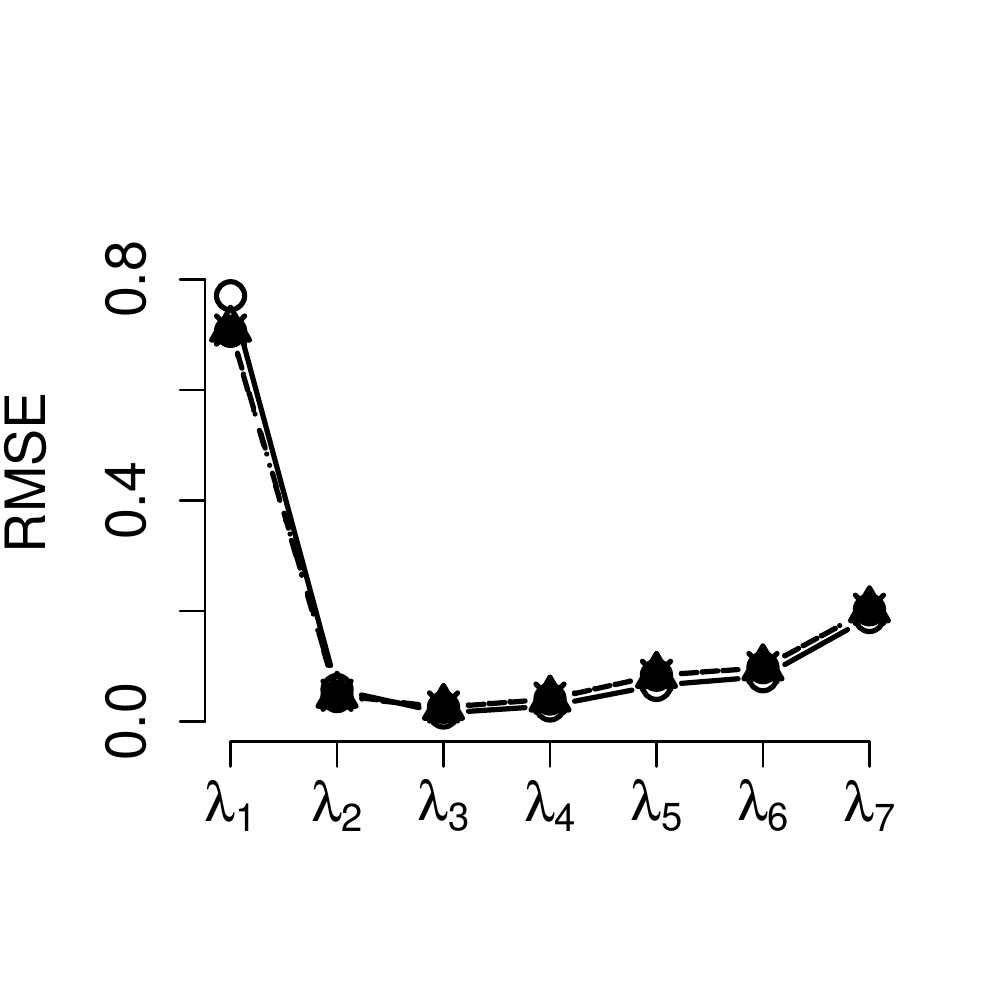}}
\end{tabular}
\vspace{-0.3 cm}\caption{{\it RMSE of each $\lambda_j$ assuming different priors for $\bflambda$. The results with the independent prior distribution and the dependent ones with $\tau=5$, $\tau=10$ and $\tau=20$ are respectively represented by the empty circles and the line, the triangles and the dashed line, the cross and the dotted line, and the full circle and the dot-dashed line.}}\label{summary_lambda_separeted}
\end{figure}

Figure \ref{summary_lambda_separeted} presents the RMSE for each $\lambda_j$ display for samples where the network $j$ is observed (a) and when it is not (b) for the four $\bflambda$ priors employed.  Figure \ref{summary_lambda_separeted} shows that the independent prior provides a smaller RMSE than the dependent one for most cases, noticeably for the smaller $\lambda_j$'s. These results do not depend heavily on the values of $\tau$. As expected, the RMSE values of the $\lambda_j$ whose network $j$ is not sampled are greater than the RMSE values of $\lambda_j$, for $j \in s$.

\begin{figure}[h!]
\begin{tabular}{c}
\hspace{-0.3 cm}{\includegraphics[scale=0.46]{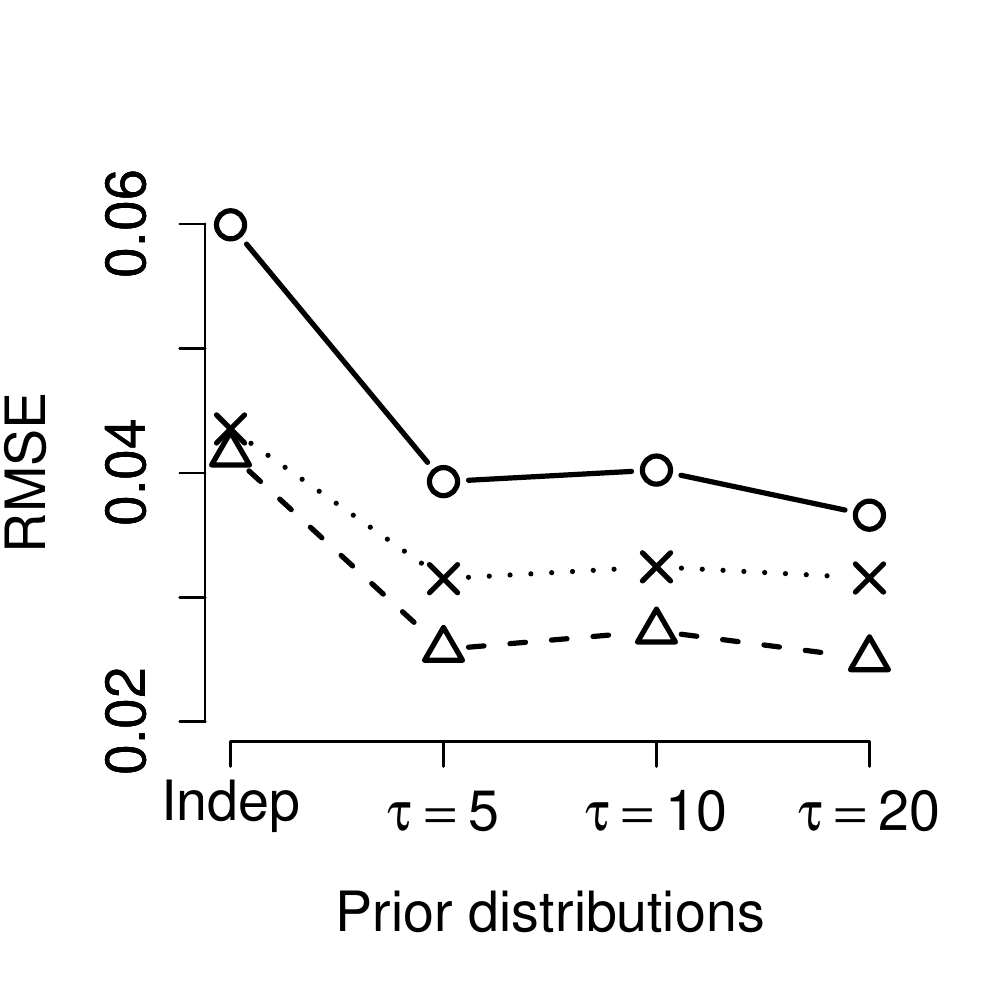}}
\hspace{-0.3 cm}{\includegraphics[scale=0.46]{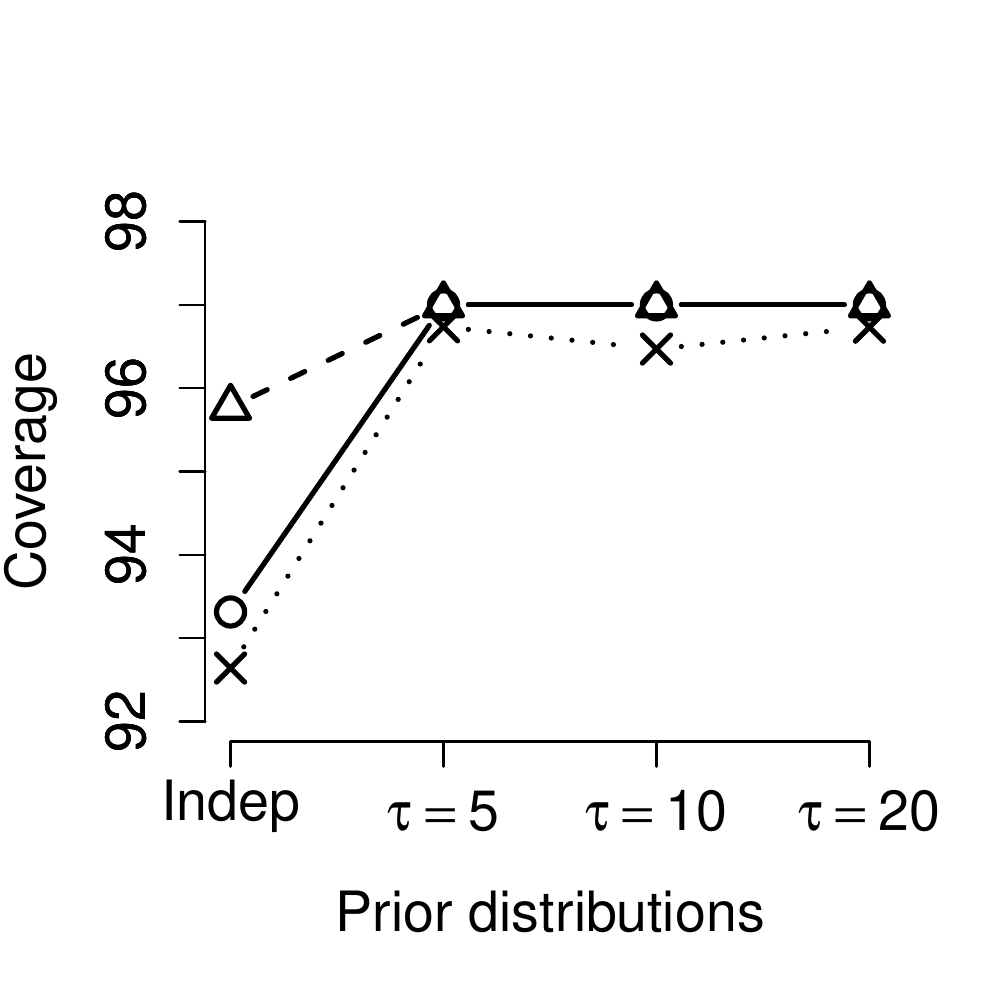}}
\hspace{-0.3 cm}{\includegraphics[scale=0.46]{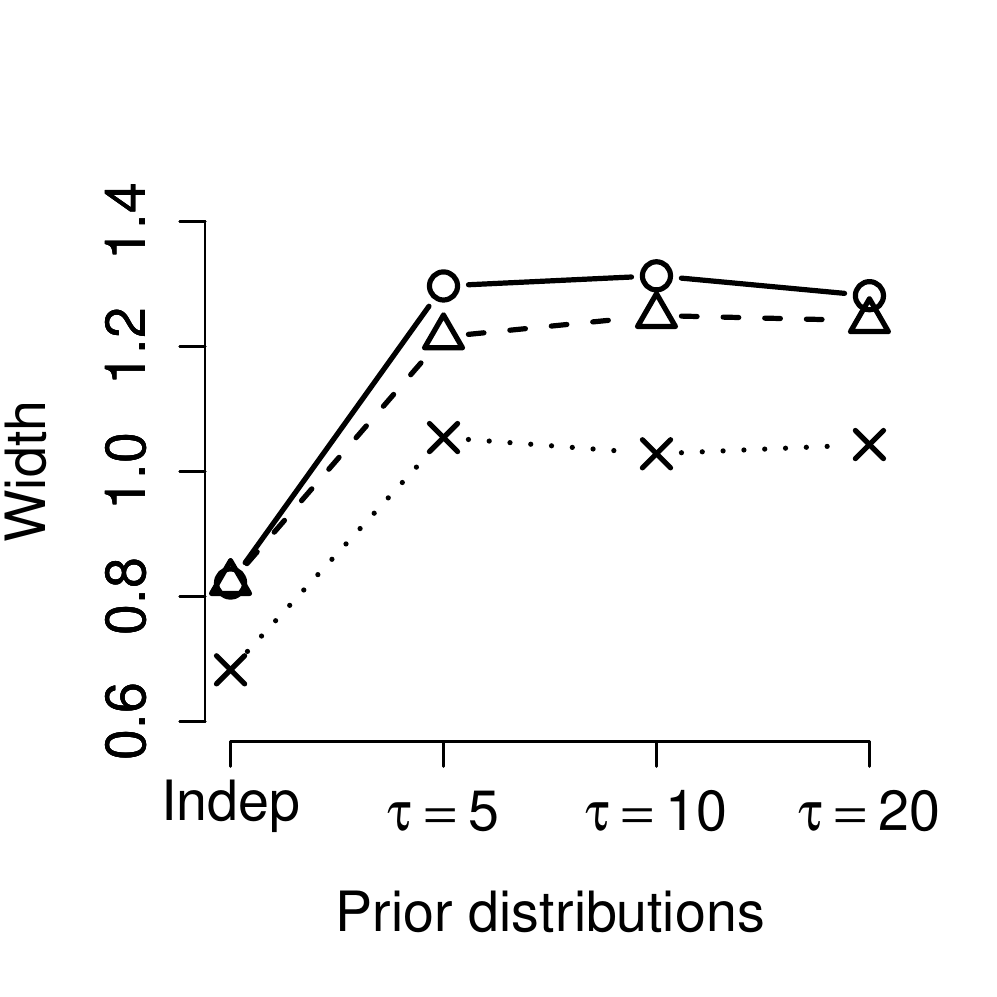}}
\end{tabular}
\vspace{-0.3 cm}\caption{{\it RMSE, coverage and widths of the 95\% HPD interval for the population total $T$ for each prior distribution assumed for $\bflambda$ and for each $R$ fixed. The results for $R=5$, $R=6$ and $R=7$ are represented by the empty circles and the line, the triangles and the dashed line, and the cross and the dotted line, respectively.}}\label{summary_T_priors}
\end{figure}

Because total population prediction is the main aim in this context, we also evaluate the impact of those prior distributions on the posterior distribution of $T$. Figure  \ref{summary_T_priors} displays the RMSE of $T$, the nominal coverage of the 95\% HPD interval and its respective width for each considered value of $R$.  We can observe from Figure  \ref{summary_T_priors} that the RMSEs obtained using the independent $\bflambda$ prior are always smaller than the ones obtained using the dependent $\bflambda$ priors. However, the 95\% HPD intervals based on the dependent $\bflambda$ priors have higher coverage than the nominal level and higher width than when using the independent $\bflambda$ prior. Note that, for a fixed value of $R$, the results provided by the dependent priors are very similar for all values of $\tau$.

\section{Comparison with the network model}\label{sec:evaluation}

The mixture model (\ref{model_mixt_prop}) has been presented as an alternative to that of  \cite{rapley2008model}.  The mixture model (\ref{model_mixt_prop}) is principally useful when we cannot assume homogeneity between networks with respect to the number of observations inside them and when the expected number of observations inside any network is not proportional to its respective area size. The key idea of this paper is to improve on the population estimates obtained by \cite{rapley2008model} through the use of a model that takes into account heterogeneity between networks. This is accomplished by modeling at the unit level rather than at the network level.

To assess the effectiveness of our methodology, we compared the results of our approach to the results obtained in \cite{rapley2008model}. The first comparison consists of a design-based experiment with a real population, and the second study is a model-based experiment. To fit both models, we assigned the same prior distributions used in Subsection \ref{dif_set}. To conduct the MCMC and RJMCMC simulations, we generated two chains of length 100,000 each, discarded the first 10,000 and then thinned the chain by taking every 90th sample value to obtain 1,000 independent samples.

\subsection{A design-based experiment}

We evaluated the proposed model (\ref{model_mixt_prop}) by performing a design-based experiment in which adaptive samples were drawn from a real, fixed population. Design-based studies are used in the context of survey sampling inference to evaluate the performance of model-based estimators under repeated samples taken from a real, fixed population where a characteristic of interest is known for all its units. This real population can be a Census or a large sample that is supposed for evaluation purposes to be the population. The main aim of this design-based experiment is to analyze the frequentist properties of the total estimators using both approaches.

\begin{figure}[h!]
\begin{center}
\vspace{-0.5 cm}
{\includegraphics[scale=0.6]{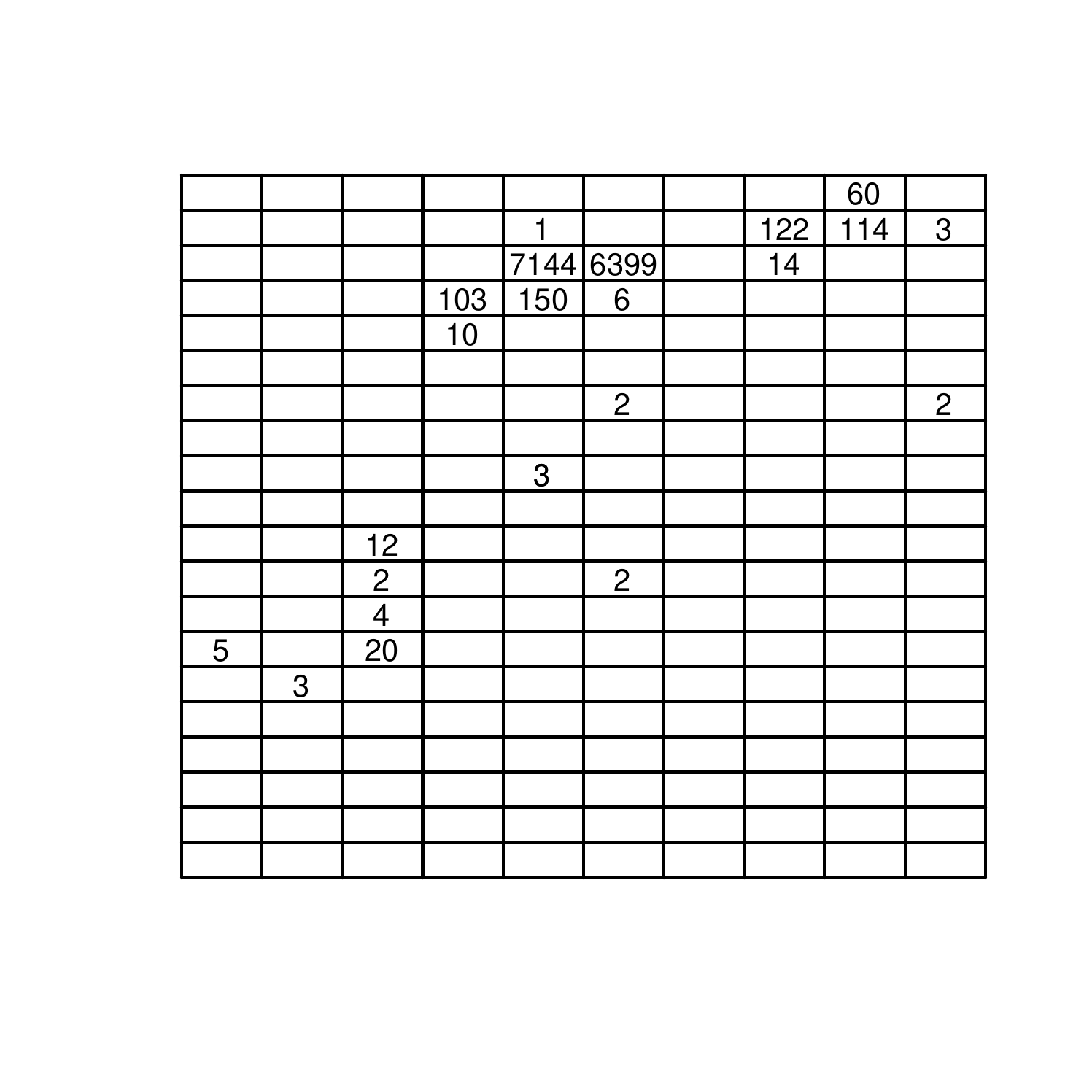}}\\
\end{center}
\vspace{-2.3 cm}\caption{{\it Counts of blue-winged teals in central Florida in 1992 in a grid with $N=200$ units.}}\label{blue_winged}
\end{figure}

 The population used here for design-based evaluation is the same described in \cite{smith1995efficiency} and consists of counts of a waterfowl species, called the blue-winged teal, in a 5,000 $km^2$ area of central Florida in 1992. Figure \ref{blue_winged} shows the counts of blue-winged teals in a grid with $N=200$ units. It should be noted that these counts are sparse and clustered, justifying the use of adaptive sampling.
 
The study consists of selecting $500$ adaptive samples with initial sizes within 10\% from the population. From now on, we will refer to the model of \cite{rapley2008model} as the '`network model'. Note that the assumptions of their model are not wholly suitable for the blue-winged teal data. Nevertheless, our proposed model assumes heterogeneity among units, which seems more reasonable when we analyze Figure \ref{blue_winged}. Furthermore, note that there are two units with a number of blue-winged teal strongly different from the others, so if the samples selected do not contain this network, it will be very difficult to accurately estimate the total population. Thus, we restricted this study to samples that contain this network, which is reasonable because the purpose is to compare the two models under the same conditions.

Figure \ref{post_dens_real_outlier} in Appendix \ref{sec:assessment} shows the trace plot with the posterior distribution of the parameters $\alpha$ and $\beta$ and the population total when fitting both models for one of the samples selected. The gray line represents the true value of the population total. Both models tend to overestimate the total, but in the network model, this error is more perceptible. The error occurs because there is one outlier network with two large values of $Y_i$. The network model is more affected by this outlier network because it assumes homogeneity within networks. The convergence was also assessed for this selected sample.  Table \ref{criteria_conv_real_outlier}  in Appendix \ref{sec:assessment} presents the values of the Geweke and Raftery-Lewis criteria. Analyzing Figure \ref{post_dens_real_outlier} and Table \ref{criteria_conv_real_outlier} leads us to conclude that convergence seems to have been reached. The same conclusion was achieved for all $500$ samples selected from this population.

 A summary comparison of the population total estimators using RMSE, RAE, and the empirical coverage of nominal 95\% HPD intervals and their widths, expressed as their respective ratios to the true values, averaged over the $500$ samples are presented in Table \ref{table_mixtmod_aplicteal}. We calculated the ratio of the variances of both Bayes estimators and referred to it as efficiency (ef) in Table \ref{table_mixtmod_aplicteal}.

Table \ref{table_mixtmod_aplicteal} shows that the network model presents larger errors than our proposed model (\ref{model_mixt_prop}).  The network model produces credible intervals that, despite their larger width, have a lower nominal coverage than desired. Furthermore, our proposed model is more efficient when applied to these data.

\begin{figure}[h!]
\begin{center}
\vspace{-0.7 cm}\begin{tabular}{cccc}
\hspace{-0.5 cm}{\includegraphics[scale=0.5]{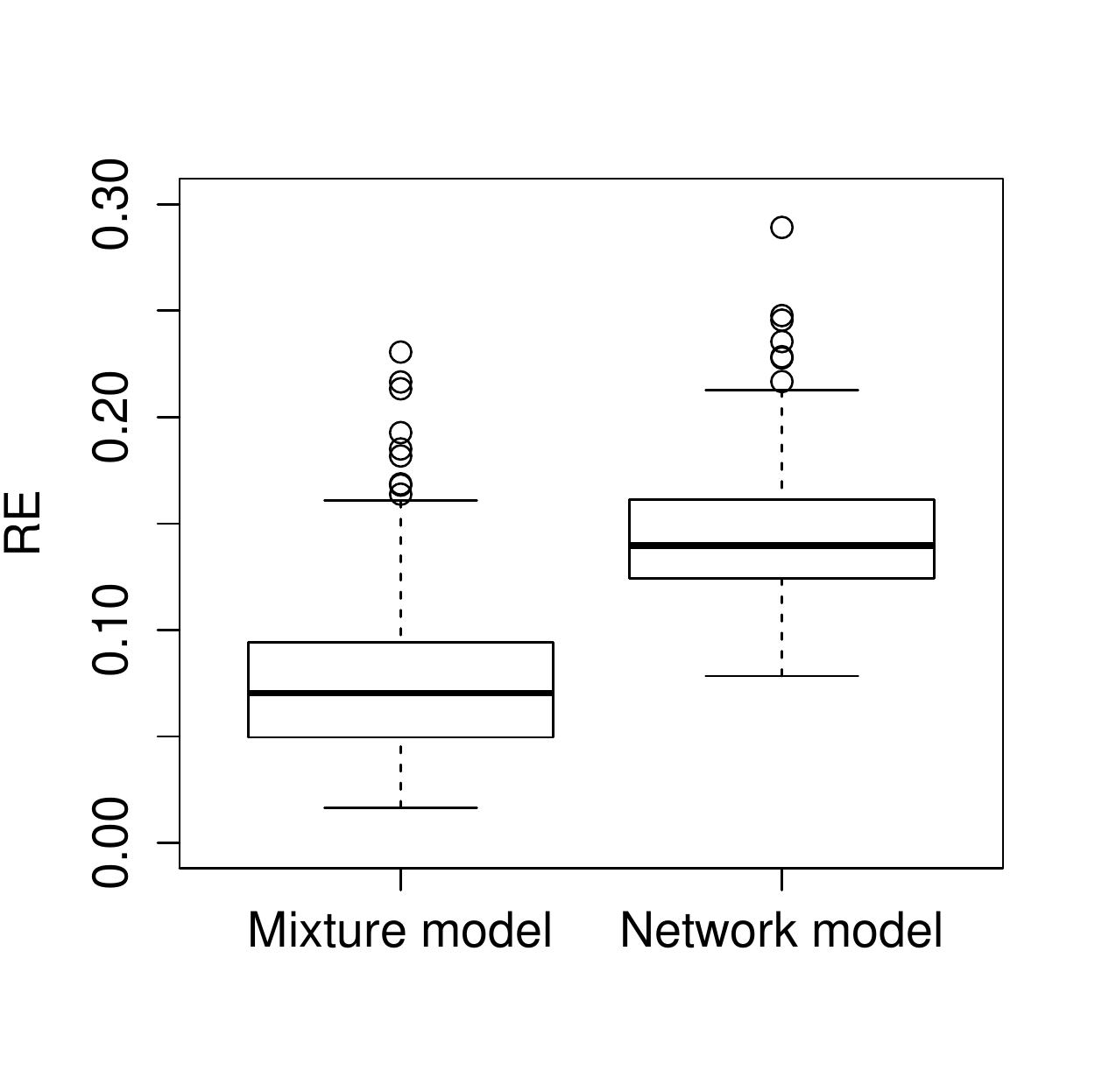}}
\end{tabular}
\end{center}
\vspace{-1 cm}\caption{{\it Boxplots with the RE for $T$ for the 500 samples, obtained by the fits of the proposed and network models.}}\label{boxplot}
\end{figure}

\begin{table*}[h]\caption{ {\it Summary measurements for the point and interval estimates of the total population, obtained by fitting the proposed and the network models.}}\vspace{0.3 cm}
	\centering
		\begin{tabular}{|c|c|c|c|c|c|c|c|c|} \hline
  & RMSE  & RAE & Coverage & Width & $\mbox{ef}(\hat{T})$ \\\hline\hline
   Mixture model & 0.01 & 0.05 & 96.7 & 0.25 & \multirow{2}{*}{0.87}\\
   Network model & 0.03 & 0.13 & 85.6 & 0.35  & \\ \hline
		\end{tabular}\label{table_mixtmod_aplicteal}
\end{table*}
Figure \ref{boxplot} shows the boxplots with the REs for the population's total posterior means and true values based on the $500$ samples when fitting both models. Here again, we see that the REs obtained for our proposed model are lower, although both overestimate the true values. This result is not unexpected, as there is a network with a substantially different number of observations from the others.

\subsection{A model-based experiment}

The purpose of this simulation study is to compare the performance of the network and mixture models when the populations are generated according to the mixture model. We considered two scenarios. For the first scenario, we used the same populations of 500 generated in the simulation study presented in Section 4.1 and fitted the network model to evaluate its performance. In particular, we considered the case where  $(\alpha,\beta)=(0.15,0.10)$. For the second scenario, we generated the components of $\bflambda$ according to a Gamma distribution with CV=25\%. Thus, it was expected that the network model performance would improve because the homogeneity degree of  $\bflambda$'s components was higher than in the first scenario (CV=50\%).

Table \ref{table_comparison_mixtmod_CV} displays some frequentist properties of the estimators obtained by fitting the network model. To facilitate the comparison, the results when fitting the mixture model with the same populations are presented in Table \ref{table_comparison_mixtmod_CV} in parentheses. Regarding the estimation of $T$, both models have equivalent performance when CV=25\%.  However, as the degree of homogeneity decreases, the mixture model performs considerably better than the network model. However, the network model exhibits better performance than the mixture model with respect to the parameter $\beta$ for both scenarios.
 \begin{table*}[h]\caption{ {\it Summary measurements for the point and interval estimates of the network model parameters over $500$ simulations where $\bflambda$ were generated from a Gamma distribution with CV=25\% and 50\%, for $N=400$ and $(\alpha,\beta)=(0.15,0.10)$.}}\vspace{0 cm}
\hspace{-1 cm}\begin{center}
{\footnotesize\begin{tabular}{c|ccc|ccccccccccc} \hline
& \multicolumn{3}{c|}{CV=25\%} & \multicolumn{3}{|c}{CV=50\%}\\\hline
& $T$ & $\alpha$ & $\beta$ & $T$ & $\alpha$ & $\beta$\\\hline
RMSE & 0.03 (0.03) & 0.05 (0.08) & 0.18 (0.41)  & 0.05 (0.03) & 0.04 (0.04) & 0.10 (0.40)\\
RAE  & 0.17 (0.14) & 0.16 (0.22) & 0.32 (0.49)  & 0.21 (0.15) & 0.19 (0.15) & 0.37 (0.50)\\
Cov. & 96.8 (96.6) & 97.1 (91.7) & 95.6 (97.5)  & 95.6 (96.5) & 98.1 (94.7) & 97.4 (97.3)\\
Wid. & 0.86 (0.70) & 0.16 (0.12) & 0.19 (0.22)  & 0.85 (0.95) & 0.16 (0.11) & 0.18 (0.23)\\\hline
		\end{tabular}}\label{table_comparison_mixtmod_CV}
		\end{center}
\end{table*}

\begin{figure}[h!]
\begin{center}
\vspace{-0.7 cm}\begin{tabular}{cccc}
\hspace{-0.5 cm}\subfigure[CV = 25\%]{\includegraphics[scale=0.5]{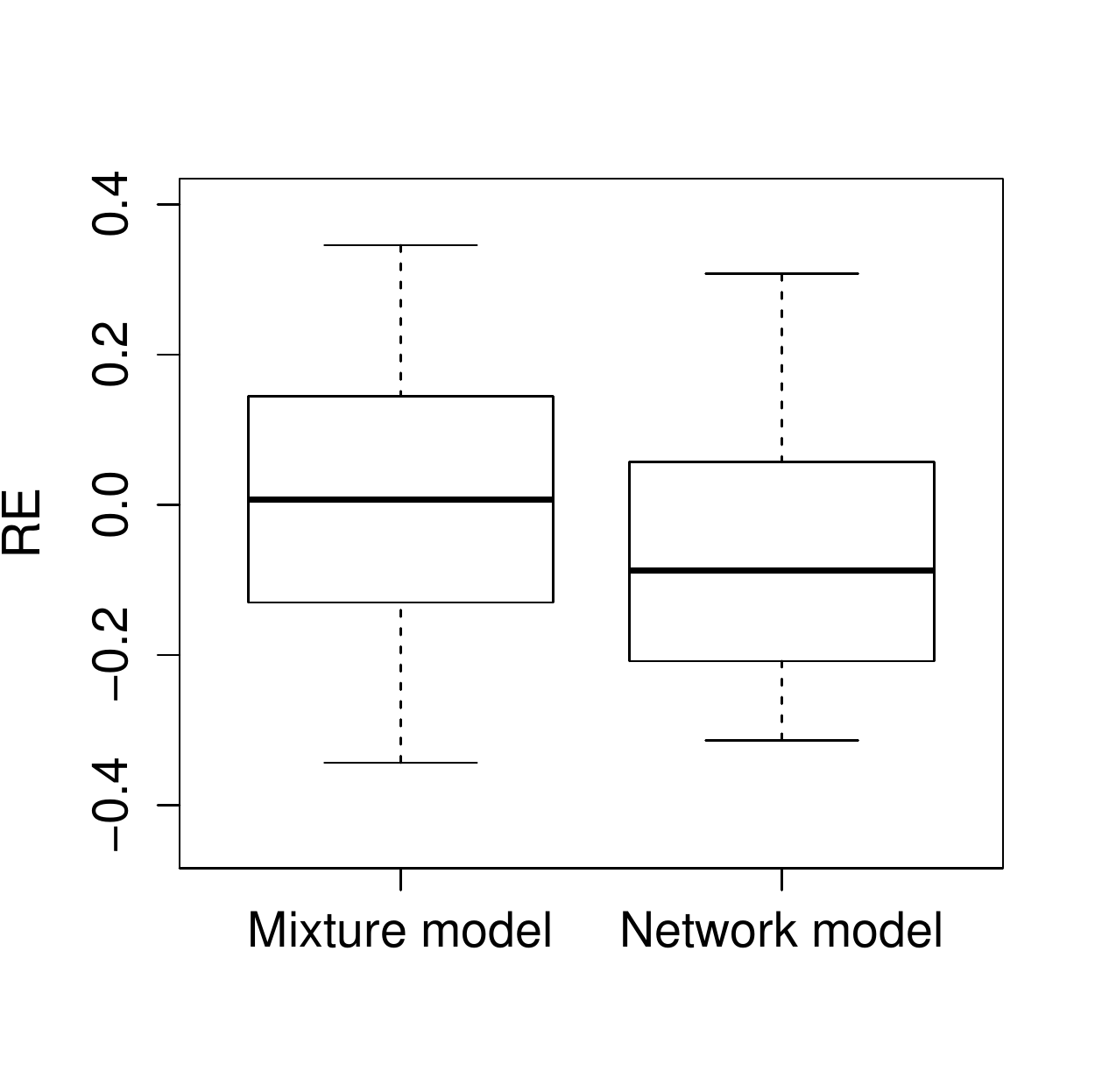}}
\subfigure[CV = 50\%]{\includegraphics[scale=0.5]{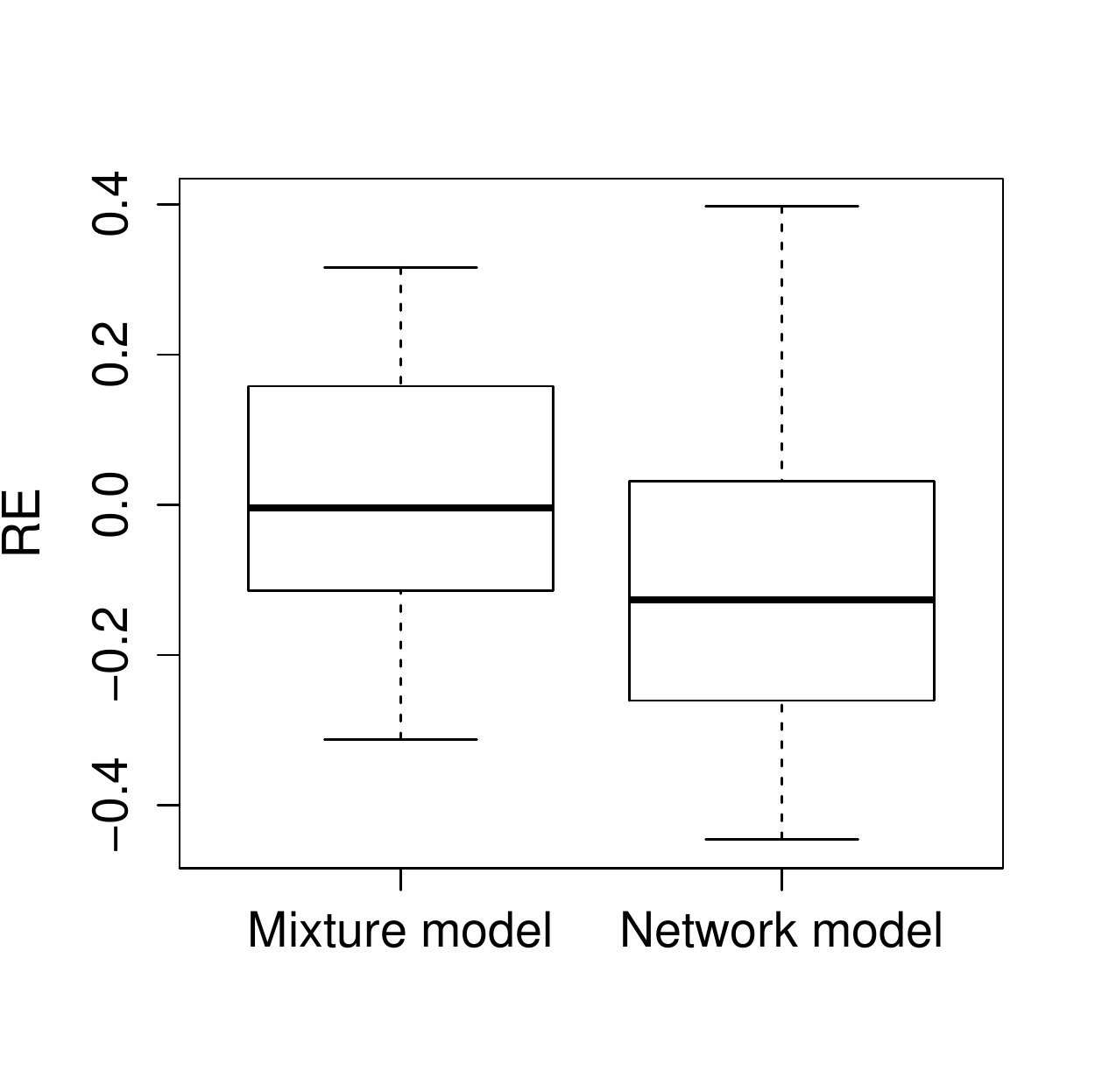}}
\end{tabular}
\end{center}
\vspace{-0.3 cm}\caption{{\it Boxplots with the RE for $T$ for the 500 samples, obtained by the fits of the proposed and the network models, for a Gamma distribution for $\bflambda$ with CV=25\% and CV=50\%.}}\label{boxplot_RET_CV25}
\end{figure}

Finally, we present the boxplot of the relative error of $T$ for both models in Figure \ref{boxplot_RET_CV25}. The conclusion is analogous to the other measurements. In particular, the estimator provided by the network model seems to underestimate $T$ for both scenarios.

Therefore, from those results, it should be concluded that as the level of homogeneity between networks increases, the performances of the evaluated models become similar. The main difference is the number of parameters to estimate and the computational effort, which is more significant when fitting the mixture model.

\section{Conclusions and suggestions for future work}\label{sec:concl}

We have considered the problem of estimating the total numbers of individuals in a rare and clustered population. Our approach is to model the observed counts in grid cells, selected by adaptive cluster sampling, and then to use model-based analysis to estimate the total population. The proposed model is an alternative to that of \cite{rapley2008model} because it models the grid cells instead of the networks and supposes heterogeneity between units that belong to different networks. Nevertheless, it requires considerable computational effort and should therefore be used only if the data support it. However, simulation studies show that as homogeneity between networks decreases, it might be worth using the mixture model as an alternative to the network model.

More general assumptions can be considered and modeled within this framework. For example, in the same network, units near the centroid should have higher frequency than units that are far from the centroid. It is possible to consider this assumption in the proposed model.

It should be noted that the parameters of the response variable associated with the unobserved components present some estimation difficulties. Therefore, the prior distribution should be carefully elicited. Thus, the main findings of this work encourage an extension of the model-based analysis to other adaptive sampling plans, which uncover more information about the population. One example is adaptive cluster double sampling, proposed by \cite{felix_adaptive_double_2004}, which allows the sampler to control the number of measurements of the variable of interest and to use auxiliary information.

{\bf Acknowledgements}

  This work is part of the Ph.D. thesis of Kelly C. M Gon\c{c}alves under the supervision of Fernando Moura, in the Graduate Program of UFRJ. Kelly has a scholarship from Coordena\c{c}\~ao de Aperfei\c{c}oamento de Pessoal do Ensino Superior (CAPES). Fernando Moura receives financial support from Conselho Nacional de Desenvolvimento Cient\'{\i}fico e Tecnol\'ogico (CNPq-Brazil, BPPesq). The authors would like to thank the editor, an associate editor and two referees for their very thoughtful and constructive comments.

\appendix
\section{Acceptance probability for the split or combination moves}\label{sec:appendix}

For the split step, to obtain the acceptance probability it is necessary to simulate $(u_1,u_2)$ from distributions with densities $g_1$ and $g_2$, respectively. The probability of acceptance, supposing an independent prior distribution for $\bflambda$, is $\min(1,A)$, where:
\begin{align*}
A&=\frac{\exp\{-(c_{j_1}\lambda_{j_1}+c_{j_2}\lambda_{j_2})\}\lambda_{j_1}^{\sum_{\{i:\epsilon_i={j_1}\}}{y_i}}\lambda_{j_2}^{\sum_{\{i:\epsilon_i={j_2}\}}{y_i}}(1-\exp(-\lambda_{j_1}))^{-c_{j_1}}
(1-\exp(-\lambda_{j_2}))^{-c_{j_2}}}{\exp\{-c_{j^*}\lambda_{j^*}\}\lambda_{j^*}^{\sum_{\{i:\epsilon_i={j^*}\}}{y_i}}(1-\exp(-\lambda_{j^*}))^{-c_{j^*}}}\\
&\frac{p(\{i_{j_1},i_{j_2}\})}{p(\{i_{j^*}\})}\times\frac{p(R_{\bar{s}}+1)}{p(R_{\bar{s}})}\times\frac{(c_{j^*}-1)!}{(c_{j_1}-1)!(c_{j_2}-1)!}(R_s+R_{\bar{s}})^{-(c_{j_1}+c_{j_2}-c_{j^*})}\times\frac{c_{j_1}^{c_{j_1}}c_{j_2}^{c_{j_2}}}{c_{j^*}^{c_{j^*}}}\times(R_{\bar{s}}+1)\\
&\times\frac{\nu^d}{\Gamma(d)}\left(\frac{\lambda_{j_1}\lambda_{j_2}}{\lambda_{j^*}}\right)^{d-1}\exp\{-\nu(\lambda_{j_1}+\lambda_{j_2}-\lambda_{j^*})\}\\
&\times \frac{p_{R_{\bar{s}}\mid R_{\bar{s}}+1}}{p_{R_{\bar{s}}+1\mid R_{\bar{s}}}P_{alloc}q(u_1)q(u_2)}\times |J|,
\end{align*}
where $p_{R_{\bar{s}+1}\mid R_{\bar{s}}}$ is the probability of choosing the split step, $P_{alloc}$ is the probability that this particular allocation is made, and $|J|$ is the Jacobean of the transformation $(w_{j^*},\lambda_{j^*}')$ to $(w_{j_1},w_{j_2},\lambda_{j_1}',\lambda_{j_2}')$.
For the corresponding combination step, the acceptance probability is $\min(1,A^{-1})$, and simple adaptations must be made because the proposal reduces the number of nonsampled networks by $1$.

\pagebreak

\section{Assessment of MCMC and RJMCMC with real data}\label{sec:assessment}

\begin{figure}[h!]
\begin{center}
\vspace{-0.7 cm}\begin{tabular}{cccc}
\hspace{-0.5 cm}{\includegraphics[scale=0.4]{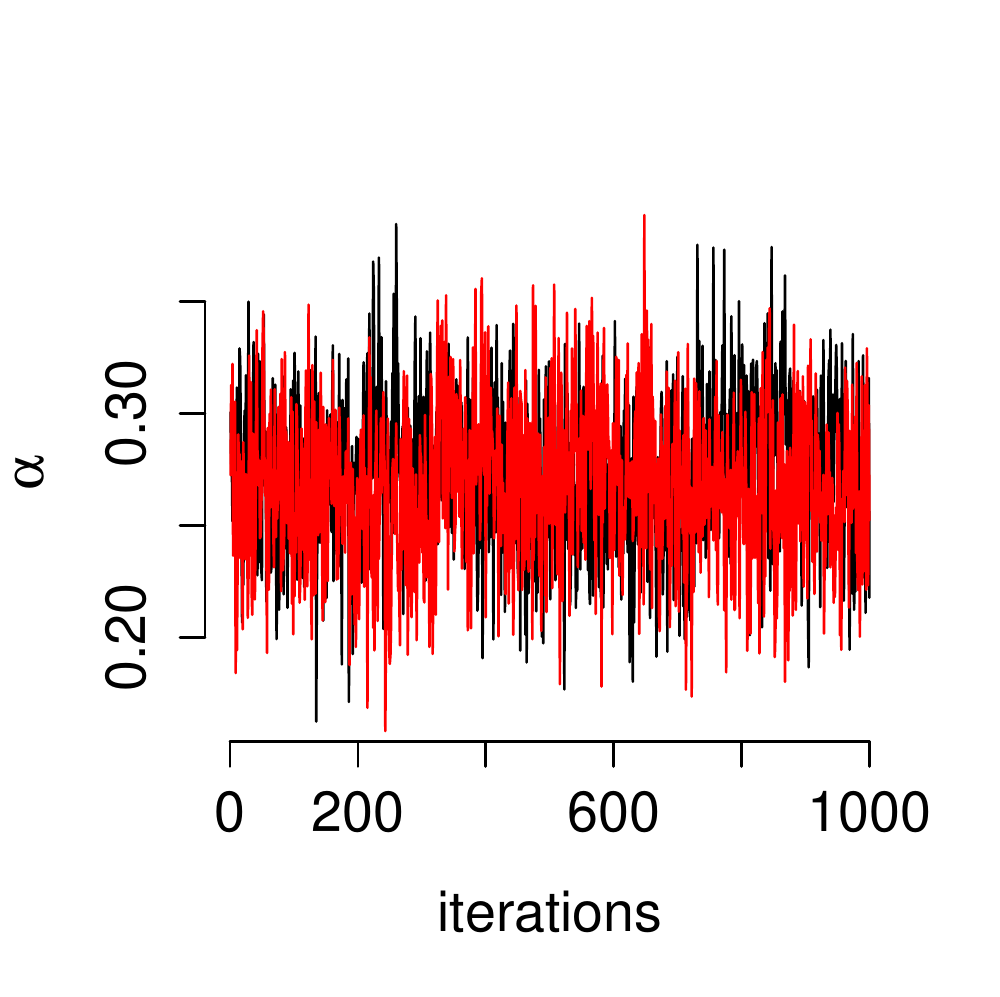}}
\hspace{-0.5 cm}\subfigure[Mixture model]{\includegraphics[scale=0.4]{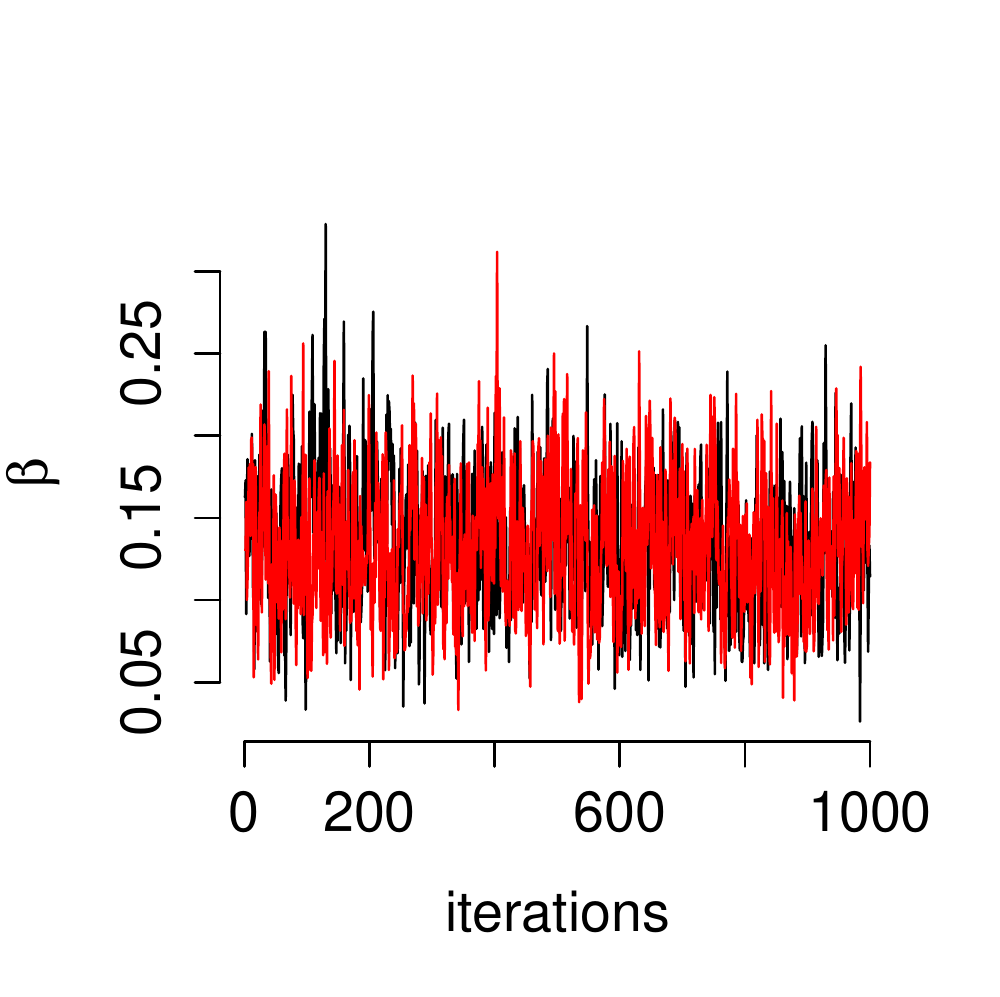}}
\hspace{-0.5 cm}{\includegraphics[scale=0.4]{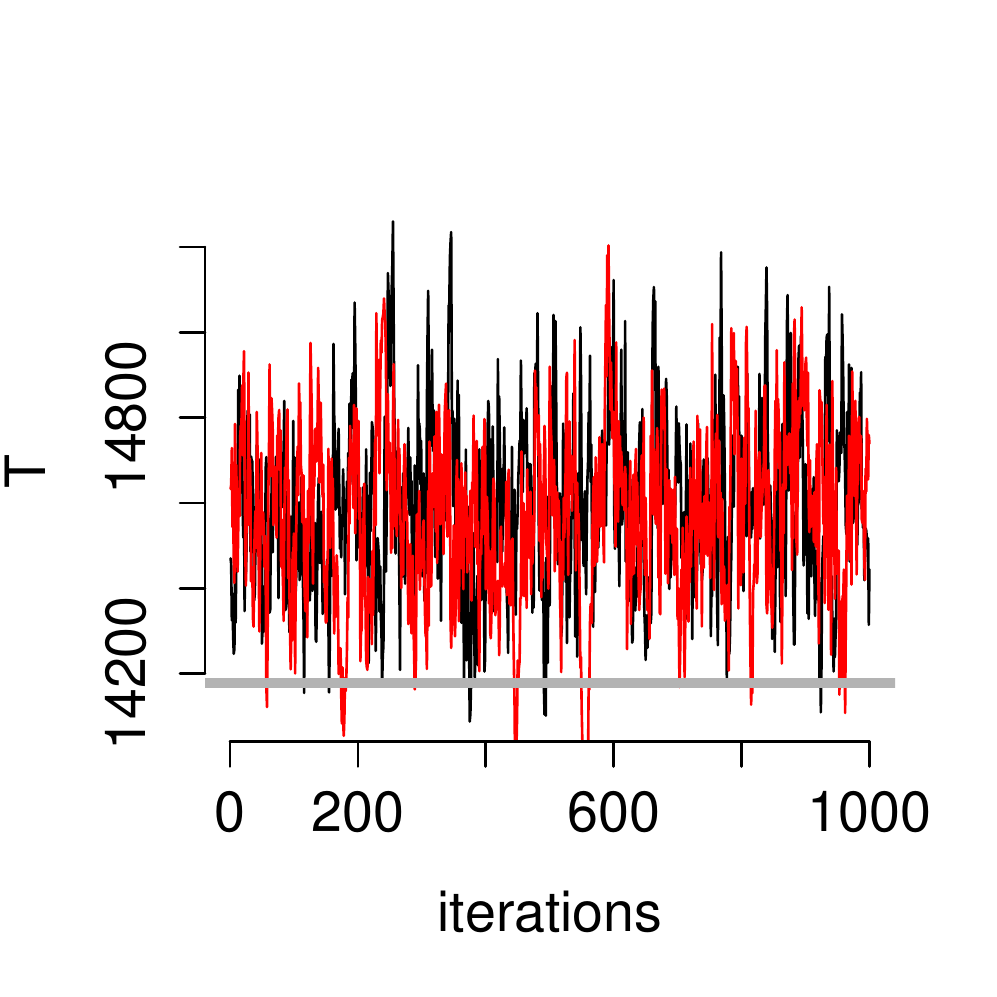}}\\
\hspace{-0.5 cm}{\includegraphics[scale=0.4]{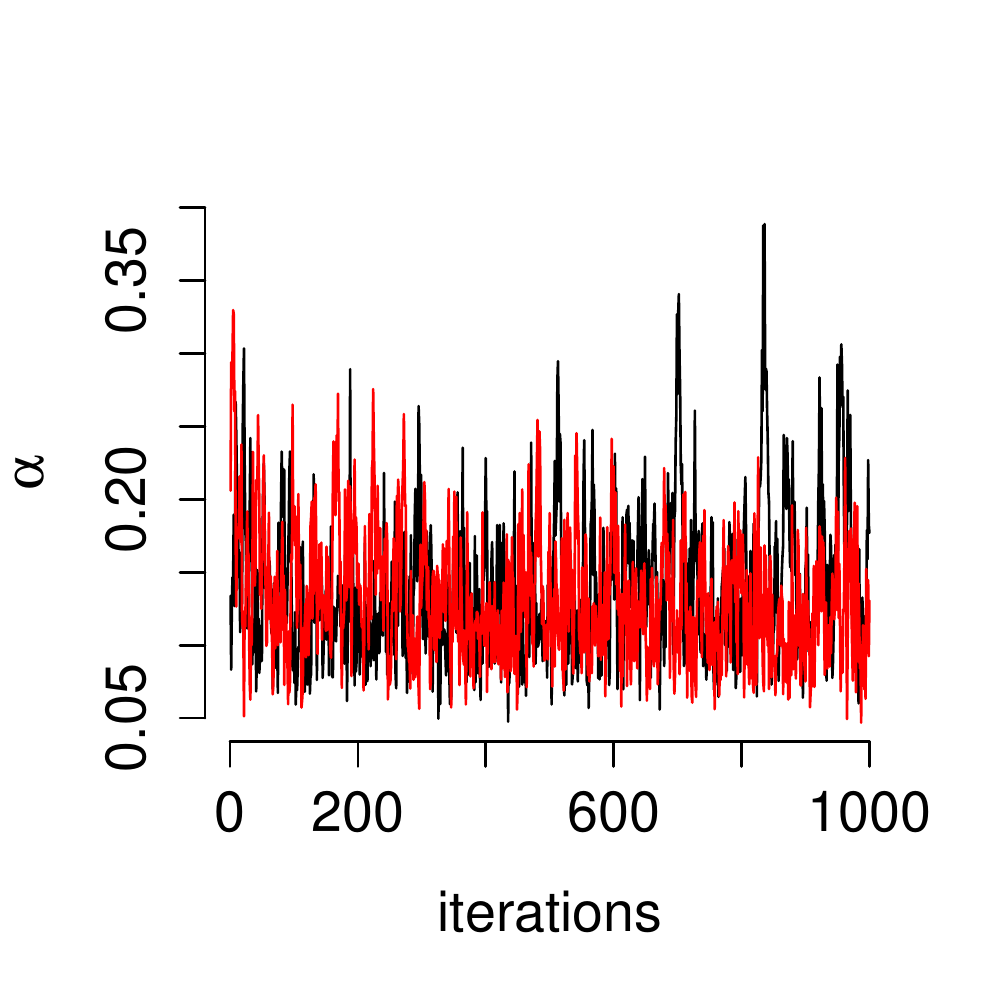}}
\hspace{-0.5 cm}\subfigure[Network model]{\includegraphics[scale=0.4]{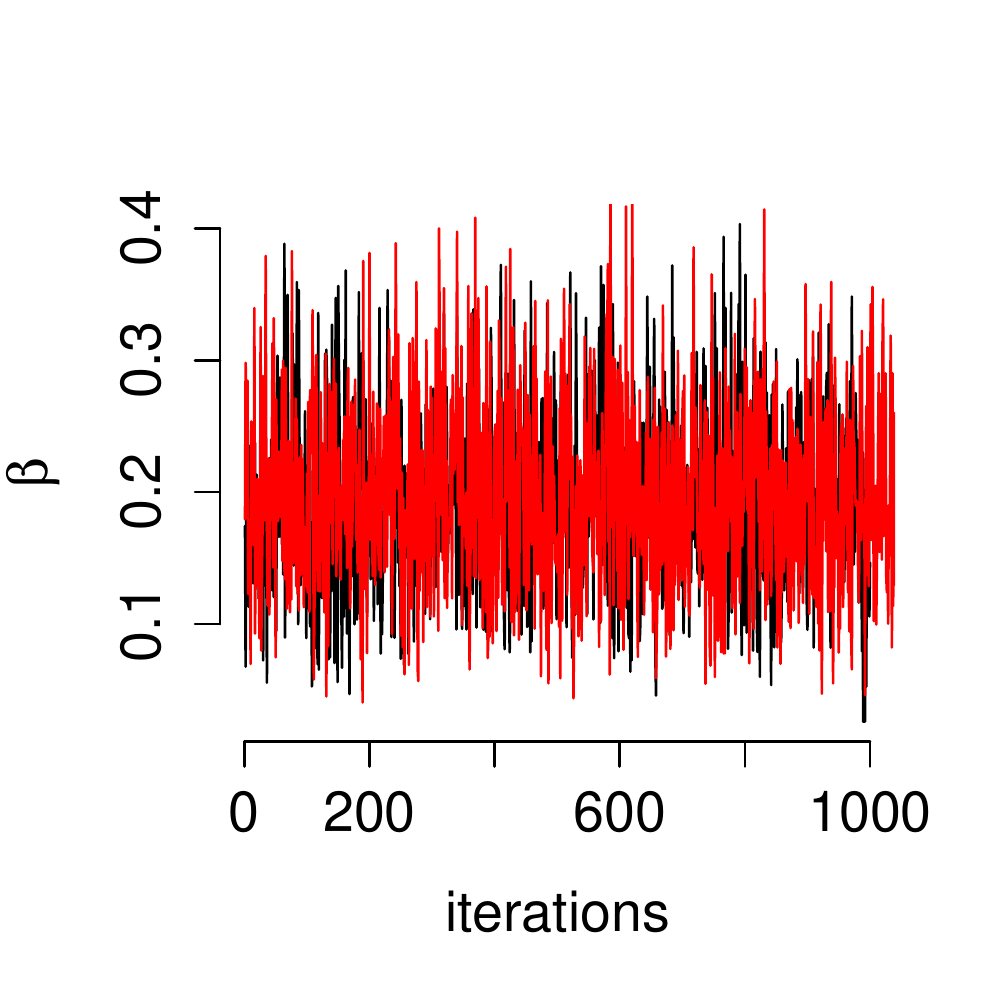}}
\hspace{-0.5 cm}{\includegraphics[scale=0.4]{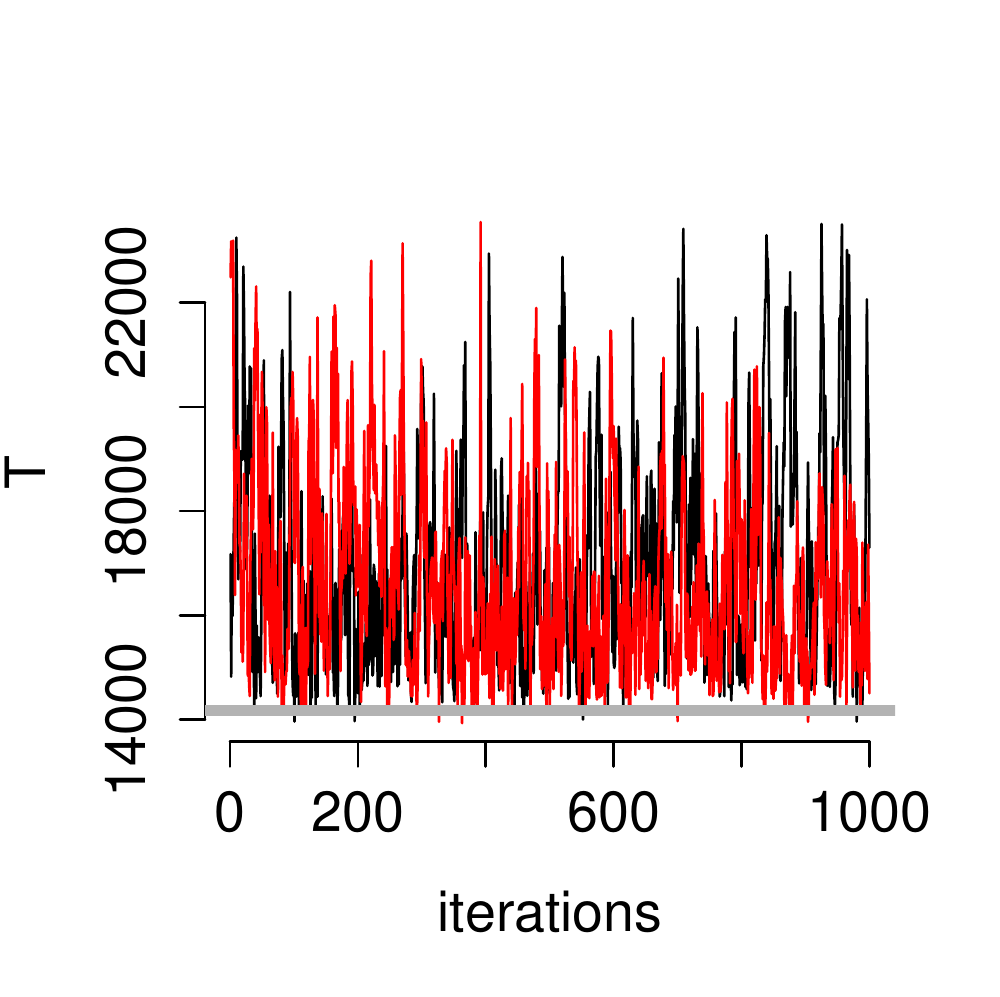}}
\end{tabular}\vspace{-0.3 cm}\caption{{\it Trace plot with the posterior densities of $\alpha$, $\beta$ and $T$ obtained by the fits of the proposed and the network models. The gray line represents the true value of $T$.}}\label{post_dens_real_outlier}
\end{center}
\end{figure}
{\footnotesize\begin{table*}[h]\caption{ {\it Geweke and Raftery-Lewis convergence diagnostics for some of the parameters estimated for the real population for both models. }}
\begin{center}
\begin{tabular}{c|cc|cccccccccccc}
& \multicolumn{2}{c}{Geweke} & \multicolumn{2}{c}{Raftery-Lewis}\\\cline{2-5}
Param & Mixture & Network & Mixture & Network\\\hline
$\alpha$ & -0.13 & -0.10 & 1.02 & 1.21\\
$\beta$ & 0.72 & -0.67  & 1.15 & 2.56 \\
$T$ & -1.38 & -0.30 & 3.22 & 1.33\\\hline
	\end{tabular}\label{criteria_conv_real_outlier}
		\end{center}
\end{table*}}

\end{document}